\documentclass[useAMS,usenatbib,fleqn]{mn2e}

\usepackage{times}

\usepackage{amsmath}
\usepackage{upgreek}
\usepackage{graphicx}

\newcommand{\pd}{\partial}                        
\renewcommand{\vec}[1]{\ensuremath{\bmath{#1}}}   
\newcommand{\hv}[1]{\ensuremath{\hat{\bmath{#1}}}}
\newcommand{\tv}[1]{\ensuremath{\tilde{\bmath{#1}}}} 
\newcommand{\mtx}[1]{\ensuremath{\mathbfss{#1}}}  
\newcommand{\vprod}{\ensuremath{\bmath{\times}}}  
\newcommand{\del}{\ensuremath{\bmath{\nabla}}}    
\newcommand{\ham}{\mbox{\ensuremath{\hat{H}}}}    
\newcommand{\td}{\mbox{\ensuremath{\tilde{\cal E}}}} 
\newcommand{\mpol}{\mbox{\ensuremath{\tilde{\cal P}}}} 

\title[OAM Radiation in Masers]{Amplification of OAM Radiation by Astrophysical Masers}
\author[M. D. Gray, G. Pisano, S. Maccalli and P.Schemmel]{M. D. Gray$^{1}$, G. Pisano$^{1}$,
S. Maccalli$^{1}$ and P. Schemmel$^{1}$\\
$^{1}$Jodrell Bank Centre for Astrophysics, School of Physics and Astronomy, University of Manchester,
M13 9PL, UK}
\begin{document}

\date{}

\pagerange{\pageref{firstpage}--\pageref{lastpage}} \pubyear{2002}

\maketitle

\label{firstpage}

\begin{abstract}
We extend the theory of astrophysical maser propagation through a medium with a
Zeeman-split molecular response to the case of a non-uniform magnetic field, and
allow a component of the electric field of the radiation in the direction of
propagation: a characteristic of radiation with orbital angular momentum. A
classical reduction of the governing equations leads to a set of nine differential
equations for the evolution of intensity-like parameters for each Fourier
component of the radiation. Four of these parameters correspond to the standard
Stokes parameters, whilst the other five represent the $z$-component of the
electric field, and its coupling to the conventional components in the $x-y$-plane.
A restricted analytical solution of the governing equations demonstrates a
non-trivial coupling of the Stokes parameters to those representing orbital
angular momentum: the $z$-component of the electric field can grow from
a background in which only Stokes-$I$ is non-zero. A numerical solution of the
governing equations reveals radiation patterns with a radial and angular structure
for the case of an ideal quadrupole magnetic field perpendicular to the
propagation direction. In this ideal case generation of radiation orbital angular
momentum, like polarization, can approach 100 per cent.
\end{abstract}


\section{Introduction}
\label{intro}

The angular momentum carried by electromagnetic radiation is not limited
to the familiar spin angular momentum, associated with polarization. It
is also possible for radiation to exhibit orbital angular
momentum (OAM) \citep{1943Phy....10..585H,1992PhRvA..45.8185A}. Like 
other forms of electromagnetic radiation, light carrying OAM
is often considered quantized into photons, and the OAM may then be referred
to as photon-OAM (POAM).
It appears to be a matter of debate as to whether the orbital and spin
angular momenta are independently quantized: work in favour 
\citep{1992PhRvA..45.8185A,2002JOptB...4S...7B} is challenged in
\citet{2008JPhB...41i5001G} and references therein, where conservation
applies to the total angular momentum and its projection on the 
propagation axis. Assuming that it is possible to make at least an
approximate separation of spin and OAM, 
rays of radiation carrying well-defined
amounts of OAM are referred to as helical modes. As with most work
based on the radio waveband, it is convenient to consider the radiation
in the form of classical fields, so from now on we will consider mostly
OAM rather than POAM. A convenient distinction may be made in the manner
of detection: POAM then refers to radiation that undergoes quantum detection,
releasing a fixed amount of energy, whilst OAM refers to radio-style
detection at the electric field level, with measurement of amplitude and
phase.

Radiation with POAM is well known in astrophysics, even if not by
this name: any quantized transition of an atom or molecule that is
not an electric dipole transition can lead to the appearance of radiation
carrying OAM: for example an
electric quadrupole transition in a molecule, with a change in
rotational quantum number of $\Delta J=\pm2$, emits a photon with two units of
angular momentum, only one of which can be spin. Such a photon is
just one example from a hierarchy of electric multipole photons, corresponding
to transitions of increasing numbers of quanta \citep{0750633719}.
An example of such a transition is the $v=1-0, S(1)$ transition of
H$_{2}$ at 2.122\,$\umu$m . As an S-branch transition, $\Delta J = 2$, and
photons emitted in this transition must carry POAM in addition to spin. This
transition has been observed from regions of shocked molecular hydrogen
at least as far back as the mid-1980s \citep{1984MNRAS.210..565G}.

In the radio region, the familiar 21-cm line of atomic hydrogen \citep{1951Natur.168..356E} must
also be a carrier of POAM because it is emitted by a magnetic dipole transition.
Although only one quantum of angular momentum is carried in this case,
POAM is required to allow the photons to have even parity. In spite of
this property of 21-cm photons, we do not expect typical H{\sc i} observations
to display an OAM signal - that is a spatial structure in the
complex amplitude of the electric field that may be mathematically represented
in terms of helical modes. Emission from most radio sources is simply too
spatially incoherent, or chaotic, for any OAM signal to be found.

The OAM properties of a helical mode propagating along the
$z$-axis, may be represented in terms of
polar coordinates, $r$, $\phi$ in the $xy$-plane. The electric
field of such a mode may be written,
\[
\bmath{E}(r,\phi ) = \bmath{E}_{0}(r) \rmn{e}^{im\phi},
\]
where $m$ is an integer, known as the helicity of the mode. More
complicated fields may be resolved into a superposition of
helical modes with a theoretically infinite range of $m$.
Consequences of the helicity of the wave-fronts of modes with
$m \neq 0$ include a Poynting vector that is not instantaneously
parallel to the axis of propagation of the radiation, and that
the electric field of the radiation includes a component parallel
to the propagation axis
(see for example, \citealt{1992PhRvA..45.8185A,2004PhT....57e..35P}). 
Radiation with OAM of
this type would not be detected by current radio telescopes due
to rapid attenuation in the detector system (for single dish instruments)
and a geographical phase offset (in interferometers).

In vacuum, or in homogeneous, isotropic media, spin and
orbital angular momenta of radiation are conserved separately
\citep{2006PhRvL..96p3905M}. Optically anisotropic media allow the exchange of
the spin angular momentum with matter, but exchange of OAM
with matter requires a transparent medium that is isotropic,
but inhomogeneous. Anisotropic, inhomogeneous media allow
OAM and spin angular momenta to be exchanged with matter
simultaneously \citep{2006PhRvL..96p3905M}. In such a medium, the helicity
of the output wave-front can be controlled by the polarization
of the input radiation.

Practical devices for generating
helical modes in the laboratory include astigmatically compensated
laser cavities \citep{1990OptCo..78..253T,2004ForPh..52.1141S}, 
lens-based mode converters \citep{1993OptCo..96..123B}, 
computer-generated holograms \citep{1992JMOp...39..985B},
spiral phase plates, for
example \citet{2004ForPh..52.1141S}, and
q-plates \citep{2006PhRvL..96p3905M}. The first of these
can be considered a source of radiation with OAM, whilst the
others convert a conventional beam into one or more helical
modes. The operation of computer generated holograms, which
resemble a diffraction grating with a fork discontinuity on the
optical axis, is
discussed in detail in \citet{2004ForPh..52.1141S}. A q-plate usually consists
of a disc of ordinary
dielectric material given birefringent properties by the incision
of a set of azimuthal grooves. For the radio or microwave region,
a disc of plastic is commonly used, for example nylon,
refractive index $n$, with a radial groove periodicity smaller
than $\lambda /2$, where $\lambda$ is the operating wavelength.
The diameter of the disc is $\gg\lambda$. Other important
parameters of the q-plate, such as the groove depth, disc thickness
and $q$ itself, the space to period ratio, are related to $\lambda$,
and controlled by formulae in \citet{1983ApPhL..42..492F}.
Parameters for a radio astronomy device of this type may be found in
\citet{2013ApOpt..52..635M}. 

In astronomy, the passage of radiation with OAM through a
variety of instruments has been considered by
\citet{2008A&A...492..883E}. In addition to free-space propagation,
Elias considers reception by an aberration-free telescope,
a coronagraph, a Michelson interferometer and a rancorimeter - a
form of correlator. The present authors are constructing
a q-plate-based detector to search for astrophysical signals
with OAM in the microwave region: its parameters have been
introduced above, and details appear in
\citet{2013ApOpt..52..635M}.
\citet{2003ApJ...597.1266H} considers 
various possibilities
for astrophysical sources of OAM-bearing radiation - one of
which is the natural maser. \citet{2003ApJ...597.1266H} suggests
that OAM is imparted by significant departures of the refractive
index from 1 within the volume of the maser - a feature of
propagation rarely considered in studies of ideal maser
amplification. A delay of order one wavelength can plausibly be
reached in a distance much smaller than a typical maser
gain length (10$^{\rmn 12}$\,cm compared with 10$^{\rmn 14}$\,cm)
at the longer maser wavelengths (for example the 1.7-GHz lines
of the OH rotational ground state). Moreover, this delay requires
only a modest ionization fraction of $\sim$10$^{\rmn -6}$ from
cosmic rays. Other possible astrophysical generators of
OAM include turbulent fields with Kolmogorov and
von Karman spectra that lead to pairs of branch points, of opposite
helicity, in a propagating electromagnetic wave \citep{2011OExpr..1925388S,2011OExpr..1924596S}.
Branch points of this type are formed by destructive interference of
an initially plane wave passing through a turbulent medium of variable refractive
index, and correspond to OAM photons \citep{2012OExpr..20.1046O}.
The branch points also correspond to locations of zero intensity in wave-front
sensors employed in the adaptive optics systems used with optical
telescopes \citep{1998JOSAA..15.2759F}. This property has enabled a real adaptive optics
system to be used as an OAM-sensitve detector in the optical regime, and
a first astronomical detection in this waveband has
been convincingly claimed by \citet{2013A&A...556A.130S}: they attempted to
detect an OAM signal via an adaptive optics system towards a sample of
five relatively nearby stars (within a few hundred pc of the Sun). A
better than 3$\sigma$ detection of OAM was obtained towards the K-type giant
HR1529 with a conversion rate of an assumed OAM-free stellar flux to OAM
on its journey to the detector of 7 per cent.

\subsection{Diagnostic Value of OAM Radiation}
\label{ss:diagnostics}

The inclusion of parameters representing OAM should provide an advance
in the diagnostic potential of radiation as great as that introduced
by considering full polarization instead of intensity. Perhaps the best that
can be said in general terms is that OAM radiation is diagnostic of
inhomogeneities in the medium through which the radiation passes, both
within an astronomical source, and in the interstellar medium. The
inhomogeneities may be in density, velocity, gravitational fields or,
in the context of the present work, magnetic fields.

As an example, consider the dispersion measure typically used in pulsar
measurements: it tells us the column density of free electrons along a
given line of sight, but we cannot tell if their distribution is smooth
or clumpy. Even with a one part per million ionization level that might
result from cosmic ray ionization, density inhomogeneitites can lead
to variations in the refractive index of the medium. These can, in turn,
lead to a 1-wavelength delay over a distance of order $10^{10}$\,m for
radiation of 20\,cm wavelength, generating OAM
\citep{2003ApJ...597.1266H}. The fraction of radiation converted to OAM,
and the spectrum of helical modes, index $m$, can tell us how clumpy the
electron distribution is - information complementary to the standard
dispersion measure. \citet{2003ApJ...597.1266H} also state that the 
mode spectrum as a function of frequency may be used to distinguish between
inhomogeneities in density, where the typical value of $m$
is proportional to $1/\nu$,
and gravitational inhomogeneities, where $m \propto \nu$. 

Velocity and density inhomogeneities are often related through
compressible turbulence, and passge of radiation through a turbulent
medium has already resulted in the first astrophysical detections of
OAM towards nearby stars
\citep{2013A&A...556A.130S,2014A&A...567A.114O}.
However, the information that can be obtained about the turbulence
from OAM radiation is impressively detailed, including the velocity 
distribution in the sky plane, and
the spatial distribution of the optical vortex pairs, resulting from
positions where the gradient of the refractive index is very large.

It may also prove possible to detect rapidly rotating objects via
OAM radiation scattered from their surfaces
\citep{2013Sci...341..537L}. This diagnostic invokes a rotational form
of the Doppler effect, and can detect rotation perpendicular to
the line of sight. A frequency shift of 
$\Delta \nu = l\Omega /(2\pi)$ appears for an object with angular velocity
$\Omega$ and radiation with $l$ units of OAM.

\subsection{Focus of the Present Work}
\label{ss:focus}

In order to represent a system more closely related to previous studies
of astrophysical maser environments, the
present work considers
the amplification of radiation with OAM by a non-uniform magnetic field,
coupled to Zeeman-split molecular energy states. We note that there is
a long history of theoretical studies of the propagation of polarized
maser radiation through a Zeeman-split molecular ensemble generated by a uniform magnetic field
(see below). Separately, non-uniform magnetic fields have been suggested
as a generator for radiation with OAM
(see for example \citealt{2003PhRvA..67b3803A}), without specific application
to masers. We note that the Zeeman effect, even in a uniform magnetic field,
renders the medium anisotropic, whilst introduction of non-uniformity makes it
also inhomogeneous: properties that allow both spin and orbital angular momentum
to be exchanged with the medium.

\subsection{Earlier Zeeman Maser Studies}
\label{ss:zhist}

One of the most striking observational phenomena associated with OH masers
in Galactic star-forming regions is their very high level (often $\sim$100
per cent) of polarization, particularly circular and elliptical polarization.
An association of these polarization properties with magnetic fields and the Zeeman effect
was suggested in the earliest days of astrophysical maser research
\citep{1965Natur.208..440W}. A theoretical description of the amplification and
saturation of polarized masers in a medium of Zeeman-split molecules from
an unsplit $J=1-0$ transition was supplied by \citet{1973ApJ...179..111G}.
This description considers several cases, depending on whether the maser is
in the limit of negligible or extreme saturation, and whether the magnetic
field is sufficiently strong to provide a good quantization axis. However, it
does not provide numerical results that show the development of polarization
through arbitrary levels of saturation.

Advances in more recent studies include generalization to transitions more
complicated than $J=1-0$ \citep{1984ApJ...285..158W,1990ApJ...354..649D} together
with numerical calculations covering arbitrary levels of saturation. The small
and large Zeeman splitting limits have been developed to arbitrary splittings \citep{1996ApJ...457..415E}
and the appearance of circular polarization as the Zeeman splitting is increased away
from zero \citep{1998ApJ...504..390E,2001ApJ...558L..55W}. Propagation of polarized
radiation through a more realistic medium, permeated by magnetohydrodynamic turbulence,
has been considered by \citet{2007ApJ...655..275W}. More accurate saturation, with
an attempt to include residual non-Gaussian statistics and coherence has been
attempted by \citet{1995A&A...298..243G} and \citet{2009MNRAS.399.1495D}.
However, as far the present authors are aware, there has been no previous attempt
to model the interaction of a molecular Zeeman system with radiation that has 
an electric field component in the direction of propagation: a necessary, but
not sufficient, condition for the presence of an OAM radiation pattern. All
that is known at present is that a uniform magnetic field can
generate polarization, under some circumstances from an unpolarized background.

The polarized maser theory papers introduced above are not particularly
consistent in the conventions they adopt with respect to the handedness of
polarization, the definitions of the Stokes parameters, and the interpretation
of the $\sigma^+$ and $\sigma^-$ labels for transitions (see Section~\ref{sss:ztrans}).
All of the above can lead to minus signs entering equations that make
interpretation difficult when comparing one theory paper with the works of
other authors. An attempt to resolve some of the conventions used has been
made by \citet{2014MNRAS.440.2988G}. In the present work we attempt to adhere
strictly to the IEEE definition regarding the handedness of polarization, the
IAU definition of Stokes-$V$, and the definition of $\sigma^\pm$ used
by \citet{1988ApJ...326..954G}.

\subsection{Styles of Amplification}
\label{ss:ampstyle}

An astrophysical maser may generate radiation with an OAM angular
pattern by two distinct processes that we will label active and passive
amplification. Only active amplification will be considered further
in the sections that follow the introduction.

In passive amplification, a thin slab of material sliced perpendicular
to the propagation direction generates changes in the set of Stokes
parameters, at each frequency in the lineshape, by interaction with the
maser molecules in the medium. Some part of these changes is then converted
into OAM via interaction with the non-uniform magnetic field only: there
is no direct interaction of OAM radiation with the molecules. Passive
amplification, if it occurs, is likely to be dominated by the saturated
parts of a maser, simply on the grounds that these have high intensity,
and there is more radiation to convert than when the maser is unsaturated.

In active amplification, we look for an interaction between the
OAM radiation itself and the maser molecules. At first we consider
only whether there can be a non-trivial coupling of the molecular
response to a component of the electric field in the propagation direction: a
component that does not exist for ordinary polarized radiation.
Active amplification can, in principle, generate OAM from an OAM-free
background, just as masers can, under the right geometrical, Zeeman
and saturation conditions, produce high degrees of polarization from
an unpolarized background.

Active amplification clearly requires radiation with OAM to induce
stimulated emission in electric dipole molecular transitions. The theory
developed below (Section~\ref{theo}) is semi-classical, with a quantum-mechanical molecular
response driven by classical fields. From the classical field point of
view, there is no problem: if a non-trivial coupling between the
molecules and the electric field of the radiation is found, then the OAM
radiation can drive the molecules. However, quantum-mechanically, it is
not obvious that OAM radiation can drive electric dipole transitions at
all: we expect, for example, that an electric quadrupole photon (which has
one unit of OAM) to be unable to interact with electric dipole transitions
with, say, $\Delta J = \pm 1$. Experimentally, however, It does appear that
radiation with OAM can be made to interact with ordinary electric dipole
transitions in the presence of a suitable non-uniform magnetic field
\citep{2003PhRvA..67b3803A}. The magnetic field imposes a geometrical
phase that allows the OAM to interact with the electric dipole moment
of an atomic transition: in the case of \citet{2003PhRvA..67b3803A}, the
795\,nm D1 line of $^{87}$Rb with $\Delta F = \pm 1$. It seems reasonable generally for photons
photons with 2 units of OAM ($l=2$) to interact with transitions that exchange
one unit of angular momentum, since the photon has a total of three units of
angular momentum: 2 of OAM and one of spin, and these can align to
yield 1 unit to be exchanged with the atomic or molecular transition.
However, the interaction is vastly weaker than 
for an $l=0$ electric dipole photon \citep{2008JPhB...41i5001G}.

\section{Theory}
\label{theo}

The theory here generalizes earlier polarized maser theory to
the case of a non-uniform magnetic field and a non-zero component of
the electric field in the propagation direction: a necessary property
of OAM modes. In particular, the theory follows \citet{1978PhRvA..17..701M}, with
additional polarization-specific detail from \citet{2009MNRAS.399.1495D}.
Although the analysis is performed very generally, we consider results
only in the limit of small signals, so that a classical reduction
in terms of Stokes parameters is accurate, and a more advanced semi-classical
saturation as used in \cite{1995A&A...298..243G} is not required. 

\subsection{Axis Systems}
\label{axes}
\begin{figure}
\includegraphics[width=84mm]{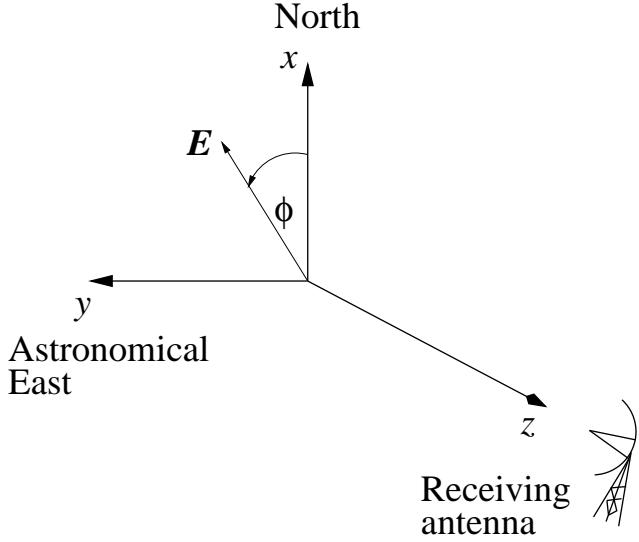}
\caption{The global axis system, conforming to the IAU definition with the
$x$-axis aligned to North and the $y$-axis to East. The electric field $\vec{E}$ is
here shown in the $xy$-plane at angle $\phi$ from the $x$-axis.}
\label{fig_iau_axes}
\end{figure}

We adopt a global, Cartesian, right-handed axis system, $(x,y,z)$, based on the propagation
direction of the radiation. Specifically, we use the standard IAU
axis system \citep{1996A&AS..117..161H} that is
drawn in Fig.~\ref{fig_iau_axes}. Radiation propagates along
the $z$-axis towards the observer, in a direction of increasing $z$.
This definition is supplemented by a set of cylindrical polar coordinates,
$(r,\phi,z)$, based on the same $z$-axis, with $r^2=x^2+y^2$ and
$\phi = \arctan (y/x)$; the angle $\phi$ is measured anticlockwise from
North.

The magnetic field, $\vec{B}$, is not uniform, so there can be no
global system of coordinates based upon the magnetic field. Quantisation 
of angular momentum along an axis based on the magnetic field can
therefore only be defined locally, that is at a specific point
$(r,\phi,z)$. It will be assumed that the magnetic field is constant
in time, but can be represented in the global coordinates as
\begin{equation}
\vec{B} = (B_x,B_y,B_z) = (B_r,B_\phi,B_z) .
\label{eq_Bdef}
\end{equation}
At any point $(x,y,z)$ we define a local set of coordinates, also
a right-handed Cartesian system, based on the local magnetic field.
This is important from the point of view of defining the correct
electric dipole alignments under the Zeeman effect (see Section~\ref{sss:dipor}).
The local system is therefore,
\[
(x',y',z') = (x'(x,y,z),y'(x,y,z),z'(x,y,z)),
\]
and is arranged at each point such that
\begin{equation}
\vec{B} = B \vec{z}' (x,y,z).
\label{eq_Bvect}
\end{equation}

\begin{figure}
\includegraphics[width=84mm]{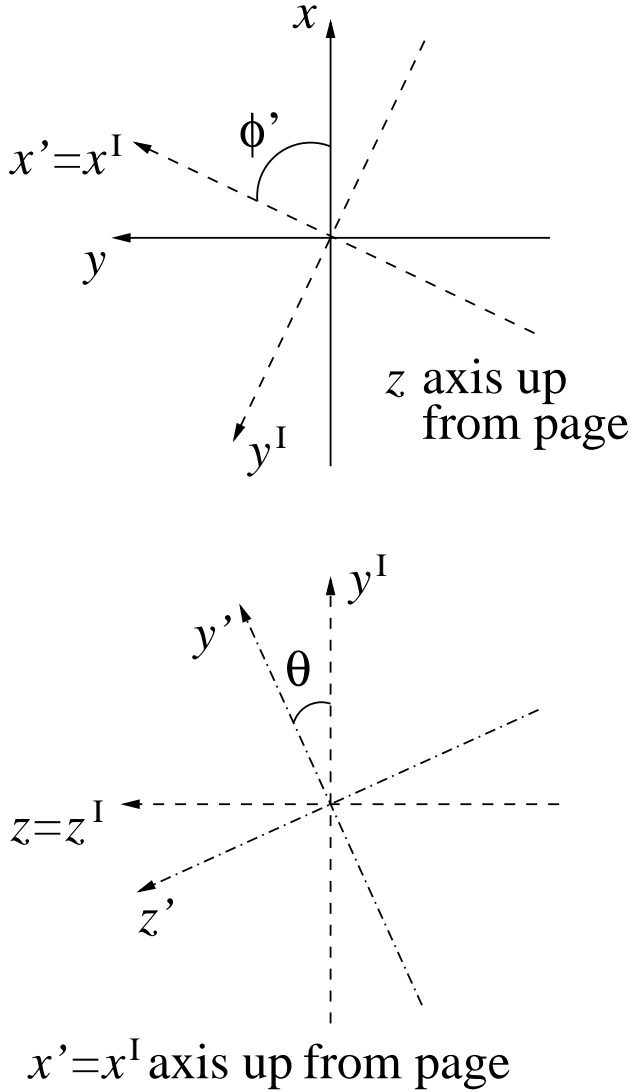}
\caption{The two rotations required to specify the components
of a vector $(x,y,z)$ in the $(x',y',z')$ system, leaving the
vector itself unaffected. The rotation in the upper part of the
figure is applied first.}
\label{fig_axrot}
\end{figure}

Without loss of generality, we can specify one local axis to
lie in the $xy$-plane, and we will choose the $x'$-axis for
this purpose. Two rotations are therefore required to represent
a vector defined in the $(x,y,z)$ system to one in the local
$(x',y',z')$ system: a rotation through an angle $\phi'$ 
about the $z$-axis to
an intermediate system $(x^I,y^I,z^I)$ in which $x^I$ is
now aligned with $x'$ and $(y^I,z^I)$ is coplanar with $(y',z')$, followed
by a rotation through $\theta$ about the $x^I$ axis
to align $(y^I,z^I)$ with $(y',z')$. Both rotations are anticlockwise
viewed in the direction of decreasing $z$ (for the first rotation),
or $x'$ (for the second). The rotations are drawn in Fig.~\ref{fig_axrot}, noting that
the vector $(x,y,z)$ remains unchanged. The matrices corresponding
to these rotations are defined in, for example,
\citet{arfken} and known here as $\mtx{R}_z(\phi')$ and $\mtx{R}_{x'}(\theta)$.
Applied sequentially, these matrices give us,
\[
\left(
\begin{array}{c}
x' \\ y' \\ z'
\end{array}
\right)
= \mtx{R}_{x'}(\theta)\mtx{R}_{z}(\phi')
\left(
\begin{array}{c}
x \\ y \\ z
\end{array}
\right) ,
\]
and we can combine these two matrices into the single
product $\mtx{R}(\theta,\phi')=\mtx{R}_{x'}(\theta)\mtx{R}_{z}(\phi')$, where
\[
\mtx{R} = \left(
   \begin{array}{ccc}
      1    &    0        &    0     \\
      0    & \cos \theta & \sin \theta \\
      0    &-\sin \theta & \cos \theta 
   \end{array}
               \right)
               \left(
   \begin{array}{ccc}
    \cos \phi'    & \sin \phi'   &    0     \\
    -\sin \phi'   & \cos \phi'   &    0     \\
      0           &      0       &    1
   \end{array}
               \right),
\label{eq_rprod}
\]
and evaluating this product, we obtain the
overall rotation matrix,
\begin{equation}
\mtx{R}(\theta,\phi') = \left(
   \begin{array}{ccc}
      \cos \phi'                 & \sin \phi'             &    0        \\
      -\cos \theta \sin \phi'    & \cos \theta \cos \phi' & \sin \theta \\
      \sin \theta \sin \phi'     &-\sin \theta \cos \phi' & \cos \theta 
   \end{array}
                        \right),
\label{eq_rotfwd}
\end{equation}
with inverse,
\begin{equation}
\mtx{R}^{-1}(\theta,\phi') = \left(
   \begin{array}{ccc}
      \cos \phi'  & -\cos \theta \sin \phi'& \sin \theta \sin \phi'  \\
      \sin \phi'  & \cos \theta \cos \phi' & -\sin \theta \cos \phi' \\
          0       & \sin \theta            & \cos \theta 
   \end{array}
                        \right),
\label{eq_rotinv}
\end{equation}
noting that $\mtx{R}^{-1}$ operates on a vector in the primed (magnetic field)
axis system, yielding its components in the unprimed (radiation) system.
If we choose, for example, the unit vector
$\hv{z}'=(0,0,1)$ and use this as the right hand side of the rotation,
\[
(x,y,z)^{\rmn T} = \mtx{R}^{-1}(\theta,\phi') (0,0,1)^{\rmn T},
\]
we find that the unit vector from the primed system has the unprimed
coordinates,
\[
\hv{z}' = \hv{z} \sin \theta \sin \phi' 
        - \hv{y} \sin \theta \cos \phi'
        + \hv{z} \cos \theta .
\]
On developing similar equations for $\hv{y}'$ and $\hv{z}'$, we
find that a vector of unit vectors in the primed system transforms
as
\begin{equation}
\left(
\begin{array}{c}
\hv{x}' \\ \hv{y}' \\ \hv{z}'
\end{array}
\right) =
\left(
   \begin{array}{ccc}
      \cos \phi'                 & \sin \phi'             &    0        \\
      -\cos \theta \sin \phi'    & \cos \theta \cos \phi' & \sin \theta \\
      \sin \theta \sin \phi'     &-\sin \theta \cos \phi' & \cos \theta 
   \end{array}
                        \right)
\left(
\begin{array}{c}
\hv{x} \\ \hv{y} \\ \hv{z}
\end{array}
\right),
\label{eq_cartgold}
\end{equation}
noting that the rotation matrix in eq.(\ref{eq_cartgold}) is $\mtx{R}$, rather than its inverse.

The angles $\theta$ and $\phi'$ may be calculated as follows:
the dot product $\vec{B} \cdot \hv{z} = B\cos \theta = B_z$, so if
$B_z(x,y,z)$ is known from the global functional form of $\vec{B}$, then
the angle is
\begin{equation}
\theta = \arccos ( B_z / \sqrt{B_x^2 + B_y^2 + B_z^2} ) = \arccos (B_z/B).
\label{eq_deftheta}
\end{equation}
A vector in the $xy$-plane corresponding to the direction of $\hv{x}'$ is
given by the cross product of $\vec{B}$ on $\hv{z}$, or
$\vec{B} \vprod \hv{z} = \hv{x} B_y - \hv{y} B_x$, and, because the magnitude
of this result is $\sqrt{B_x^2 + B_y^2}$, the local $\hv{x}'$ vector is,
\begin{equation}
\hv{x}' = (\hv{x} B_y - \hv{y} B_x)/\sqrt{B_x^2 + B_y^2}.
\label{eq_xhatfromB}
\end{equation}
The angle $\phi'$ is the offset between the $\hv{x}$ and $\hv{x}'$ unit
vectors, and may therfore be calculated by dotting $\hv{x}$ onto
eq.(\ref{eq_xhatfromB}) and setting the result equal to $\cos \phi'$.
The angle is recovered as,
\begin{equation}
\phi' = \arccos (B_y/\sqrt{B_x^2 + B_y^2}).
\label{eq_defphiprime}
\end{equation}

\subsubsection{Spherical Vector System}
\label{sss:svect}
We have obtained a transformation of the Cartesian unit vectors as
eq.(\ref{eq_cartgold}). We now wish to re-write this transformation in terms
of the spherical basis vectors that may be used to represent radiation
polarization. We define these here, in terms of the Cartesian unit vectors, as 
\begin{subequations} \label{eq_rldef}
\begin{align}
\hv{e}_R &= (\hv{x} + i \hv{y}) / \sqrt{2} \label{eq_rldef_R}\\
\hv{e}_L &= (\hv{x} - i \hv{y}) / \sqrt{2} \label{eq_rldef_L},
\end{align}
\label{eq_rldef}
\end{subequations}
a form that follows the definitions in \citet{1973ApJ...179..111G}. The
unit vector $\hv{z}$ is common to both the spherical and Cartesian systems.
In the local primed frame, exactly the same relationships apply, so that
\begin{subequations} \label{eq_rldefprime}
\begin{align}
\hv{e}'_R &= (\hv{x}' + i \hv{y}') / \sqrt{2} \label{eq_prime_R}\\
\hv{e}'_L &= (\hv{x}' - i \hv{y}') / \sqrt{2} \label{eq_prime_L}.
\end{align}
\end{subequations}
The definitions in eq.(\ref{eq_rldef}) may be inverted, yielding the Cartesian
unit vectors,
\begin{subequations} \label{eq_cart}
\begin{align}
\hv{x} &= (\hv{e}_R + \hv{e}_L ) / \sqrt{2} \label{eq_cart_x}\\
\hv{y} &= -i(\hv{e}_R - \hv{e}_L ) / \sqrt{2} \label{eq_cart_y},
\end{align}
\end{subequations}
and similarly for the primed system. When we use eq.(\ref{eq_cart}), and its
primed counterpart,
to eliminate $\hv{x}$ and $\hv{y}$ from eq.(\ref{eq_cartgold}), we obtain the
analogous expression in the spherical system
\newcounter{mytempeqncnt}
\begin{figure*}
\normalsize
\setcounter{mytempeqncnt}{\value{equation}}
\setcounter{equation}{11}
\begin{equation}
\left(
\begin{array}{c}
\hv{e}_R' \\ \hv{z}' \\ \hv{e}_L'
\end{array}
\right) =
\left(
   \begin{array}{ccc}
(1+\cos \theta)(\cos \phi' -i\sin \phi')/2    & (i\sin \theta )/\sqrt{2} & (1-\cos \theta)(\cos \phi' +i\sin \phi')/2\\
\sin \theta (\sin \phi' + i\cos \phi')/\sqrt{2}& \cos \theta              & \sin \theta (\sin \phi' - i\cos \phi')     \\
(1-\cos \theta)(\cos \phi' -i\sin \phi')/2    & (-i\sin \theta)/\sqrt{2} & (1+\cos \theta)(\cos \phi' +i\sin \phi')/2 
   \end{array}
                        \right)
\left(
\begin{array}{c}
\hv{e}_R \\ \hv{z} \\ \hv{e}_L
\end{array}
\right).
\label{eq_sphgold}
\end{equation}
\setcounter{equation}{\value{mytempeqncnt}}
\addtocounter{equation}{1}
\hrulefill
\vspace*{4pt}
\end{figure*}
This last equation is set as a two-column equation and may appear displaced from here.

\subsection{The Electric field}
\label{ss:efield}
We have noted in the Introduction that a distinguishing feature of
radiation that carries OAM is a non-zero component of the electric field
in the direction of propagation. We therefore write a general analytic
signal for the radiation that includes all three Cartesian components:
\begin{equation}
\tv{E}(\vec{r},t) = \hv{x} {\tilde E}_x(\vec{r},t)
                  + \hv{y} {\tilde E}_y(\vec{r},t)
                  + \hv{z} {\tilde E}_z(\vec{r},t),
\label{eq_sig0}
\end{equation}
where the tilde symbol over an electric field component indicates a
complex-valued quantity. For this section, and that on
radiative transfer,  we use the general
position vector $\vec{r}$, and only adopt problem-specific coordinates
in Section~\ref{sss:simp}.
The real-valued electric field is simply the
real part of the analytic signal,
\begin{equation}
\vec{E}(\vec{r},t) = \Re \{ \tv{E}(\vec{r},t) \}.
\label{eq_efield0}
\end{equation}

The electric field that we consider has a total spectral width $\Delta \nu$ that
is narrow in the sense that $\Delta \nu \ll \nu_0$, where $\nu_0$ is some representative
frequency within the band of width $\Delta \nu$. However, the radiation is broad-band
in the sense that it is inhomgeneously broadened by the Doppler effect, such that
$\Delta \nu$ vastly exceeds any frequency range directly determined by the
molecular response (the homogeneous width). To deal with this, we extract from
eq.(\ref{eq_sig0}) a rapidly oscillating term,
\begin{equation}
e^{-iY_0(\vec{r},t)} = e^{-i\omega_0 (t - \hv{n}\cdot \vec{r}/c)} ,
\label{eq_Y0}
\end{equation}
where $\omega_0 = 2\pi \nu_0$, and $\hv{n}$ is a unit vector in the direction
of radiation propagation. To obtain the last form in eq.(\ref{eq_Y0}), we adopt
the paraxial approximation of negligible beam divergence. In this case, even
for radiation with OAM, 
the direction of propagation lies along a single axis, taken here to be the $z$ axis.
The instantaeous Poynting vector, however, is not aligned with $\hv{z}$ \citep{2011allen_padgett}.
This extraction operation leaves eq.(\ref{eq_sig0}) in the form,
\begin{equation}
\tv{E}(\vec{r},t) = [\hv{x} \td_x(\vec{r},t)
                  + \hv{y}  \td_y(\vec{r},t)
                  + \hv{z}  \td_z(\vec{r},t) ] e^{-iY_0},
\label{eq_sig1}
\end{equation}
where $\td_q(\vec{r},t)$, for $q=x,y,z$, is a complex amplitude that now
has only a slow variation in $\vec{r}$ and $t$ in the sense that
$\pd \td_q /\pd t \ll \omega_0 \td_q$ and
$|\vec{\nabla} \td_q| \ll \omega_0 \td_q /c$. 

Note that the choice of the
sign $e^{-iY_0}$ (rather than $e^{+iY_0}$) in eq.(\ref{eq_sig1})
ensures that the spherical unit
vectors from eq.(\ref{eq_rldef}) correctly represent right- and left-handed
circular polarization under the IEEE convention \citep{1996A&AS..117..161H}.
We will often use a representation of the electric field in terms of
the spherical unit vectors, writing
\begin{equation}
\tv{E}(\vec{r},t) = [\hv{e}_R \td_R(\vec{r},t)
                  + \hv{e}_L  \td_L(\vec{r},t)
                  + \hv{z}  \td_z(\vec{r},t) ] e^{-iY_0},
\label{eq_sig2}
\end{equation}
where the right- and left-handed complex amplitudes are found to be,
\begin{subequations} \label{eq_rlamp}
\begin{align}
\td_R &= (\td_x - i \td_y) / \sqrt{2} \label{eq_rlamp_R}\\
\td_L &= (\td_x + i \td_y) / \sqrt{2} \label{eq_rlamp_L}.
\end{align}
\end{subequations}

\subsubsection{Other OAM Representations}
\label{sss:OAMother}
The representation of the electric field in eq.(\ref{eq_sig1}) is possibly
unusual in studies of OAM radiation. It seems customary to adopt a
representation, at least for laboratory studies, in terms of the
Laguerre-Gaussian modes (L-G modes). These modes themselves are used
to describe the electric and magnetic fields of the radiation perpendicular
to the $z$ axis, whilst the $z$-component of the field is proportional
to $x$ and $y$ gradients of the L-G modes.

Our electric field must satisfy the wave-equation,
\begin{equation}
\nabla^2 \vec{E} = \frac{\mu_r \epsilon_r (\vec{r})}{c^2} \frac{\pd^2 \vec{E}}{\pd t^2} ,
\label{eq_maxwav}
\end{equation}
and we show below in Section~\ref{ss:rt} that the form in eq.(\ref{eq_sig1}) does.
Various approximations have been made in obtaining eq.(\ref{eq_maxwav}), notably a
negligible conductivity in the medium, a linear (dielectric) response of the medium
to the electric field of the radiation, constant charge density and constant value
of $\mu_r$, the relative permeability. However, $\mu_r$ could be a tensor constant,
and $\epsilon_r$ could be a tensor quantity that is dependent on position (but not time).
Equation~\ref{eq_maxwav} is separable into time and space parts by the multiplicative
substitution,
\begin{equation}
\tv{E}(\vec{r},t) = \tv{E}_{\vec{r}}(\vec{r}) \tv{E}_t(t).
\label{eq_multsep}
\end{equation}
The time part becomes a wave equation, whilst the spatial part reduces to a
form of the Helmholtz equation. If the constant of separation is $-k^2$, where
$\omega = kc$, then we can write one solution of the time part, corresponding
to our choice of fast term in the analytic signal as,
\begin{equation}
\tv{E}_t(t) = A e^{-i \omega t} = A e^{-i \varpi t} e^{-i \omega_0 t} ,
\end{equation}
where $\varpi = \omega - \omega_0 << \omega$ is a small frequency within
the spectral width $\Delta \nu$. We can further combine the constant amplitude $A$
with the slowly-varying term as a complex amplitude,
\begin{equation}
{\tilde A}(t) = A e^{-i \varpi t} .
\end{equation}
The important point about this separation of variables is that we can write
the electric field from eq.(\ref{eq_multsep}) as
\begin{equation}
\tv{E}(\vec{r},t) = [\hv{e}_R \td_R(\vec{r})
                  + \hv{e}_L  \td_L(\vec{r})
                  + \hv{z}  \td_z(\vec{r}) ] e^{i\omega_0 z/c}{\tilde A}(t) e^{-i\omega_0 t},
\label{eq_sig3}
\end{equation}
where the spatial part must satisfy a Helmholtz equation. 
If the relative permeability and permittivity are constant, the paraxial
form of the Helmholtz equation 
may be solved via the LG modes, or any other suitable expansion for the
spatial part of the electric field..
We adopt instead the
electric field representation in eq.(\ref{eq_sig1}) because the relative
permittivity in the current problem is not constant, and we solve the
resulting radiative transfer problem in Section~\ref{ss:rt}. The problem
for which the LG modes are a solution may be considered a limit of
this problem for free space or a homogeneous medium.  

Another possibly useful field expansion is one in terms of the spherical harmonic
vectors, $\vec{Y}_{J,M}^{(e)}$ and $\vec{Y}_{J,M}^{(m)}$, corresponding to 
electric (e) and magnetic (m) multipole photons of total angular momentum 
quantum number $J$ and projection on the propagation axis $M$. The spherical
harmonic vectors may be resolved into components along the
propagation axis and perpendicular to it, the latter again resolved into components
following $\hv{e}_R$ and
$\hv{e}_L$ \citep{0750633719}. We note that the definitions of
the spherical unit vectors differ by multiplicative constants from those
used in the present work. OAM is present in all of the multipole photons
except the electric dipole type.

\subsubsection{Fourier Representation}
\label{sss:fourier}

The complex amplitudes in eq.(\ref{eq_sig1}) and eq.(\ref{eq_sig2}) are functions of
position and time, but they can be considered as being constructed from all the
frequencies within the spectral bandwidth. For a signal of infinite duration, we
would integrate over a continuum of frequencies, corresponding to an idealized case
of Fourier components of infinitessimal width. Astrophysical signals are limited
in time by a sampling process at the telescope, so the Fourier components in a
practical signal have a finite width of order $\delta \nu = 1/T$, where $T$ is the
sample duration.

The standard Fourier transform operations need some modification to work with
a limited time range and Fourier components of finite width in the frequency
domain. We adopt the transforms used by \citet{1978PhRvA..17..701M}, and subsequently
used by \citet{2009MNRAS.399.1495D} and \citet{mybook}. The inverse transform, from
frequency to time, becomes a sum over finite-width frequency strips, and a complex amplitude
for Cartesian or spherical spatial component $q$ is
\begin{equation}
\td_q(\vec{r},t) = (2\pi)^{-1} \sum_{n=-\infty}^\infty \td_q(\vec{r},\varpi_n) e^{-i\varpi_n (t-\hv{n}\cdot \vec{r}/c)} ,
\label{eq_ffwd}
\end{equation}
where $\varpi_n = \omega_n - \omega_0$ is a local frequency of magnitude $\ll \omega_0$, corresponding
to the centre of Fourier component $n$. Although the sum over the Fourier components has been formally
written with infinite limits, the number of strips required to cover a certain number of inhomogeneous
line widths, for example, would be a finite number. The forward transform, the inverse of that in eq.(\ref{eq_ffwd}),
transforming from the time domain to local frequency is 
\begin{equation}
\td_q(\vec{r},\varpi_n) = T^{-1} \int_{-T/2}^{T/2} \td_q(\vec{r},t) e^{i\varpi_n (t-\hv{n}\cdot \vec{r}/c)} .
\label{eq_finv}
\end{equation}
These transformations will be used extensively in Section~\ref{s:fdom}.

\subsection{Radiative Transfer}
\label{ss:rt}

The electric field introduced in Section~\ref{ss:efield} must satisfy the wave equation
\begin{equation}
\nabla^2 \vec{E} = \frac{\mu_0 \epsilon_r(\vec{r})}{c^2} \frac{\pd^2 \vec{E}}{\pd t^2} ,
\label{eq_wav0}
\end{equation},
that follows from eq.(\ref{eq_maxwav}), but 
we have now assumed a relative permeability of $1$.
The relative permittivity $\epsilon_r$ may be a scalar or tensor quantity. We
assume that the propagation medium is dielectric, so that the macroscopic polarization
of the medium, $\vec{P}(\vec{r},t)$, is linearly related to the electric field via the
formula
\begin{equation}
\vec{P}(\vec{r},t) = \epsilon_0 (\epsilon_r(\vec{r}) - 1) \vec{E}(\vec{r},t) .
\label{eq_dielec}
\end{equation}
Equation~\ref{eq_dielec} may be used to eliminate the permittivity from eq.(\ref{eq_wav0}), leaving
a wave equation in terms of $\vec{P}$:
\begin{equation}
\nabla^2 \vec{E} = \frac{1}{c^2} \frac{\pd^2 \vec{E}}{\pd t^2} + \mu_0 \frac{\pd^2 \vec{P}}{\pd t^2} .
\label{eq_wav1}
\end{equation}
We now substitute the electric field, in the anlytic signal representation of
eq.(\ref{eq_sig0}), into eq.(\ref{eq_wav1}), assuming that the
same representation can be used for $\vec{P}$. It immediately breaks down into three
scalar equations for the Cartesian components of the field. If each Cartesian
component is then put into the form used in eq.(\ref{eq_sig1}), the
various derivatives can be calculated and substituted into each scalar
wave equation. After some algebra, the details of which may be found
in \citet{mybook}, the rapidly oscillating terms are lost, and we find
that the complex amplitude of Cartesian component $q$ of the electric field
is transferred according to,
\begin{equation}
\left( \frac{\pd}{\pd t} + c \hv{n} \cdot \del \right) \td_q = 
   \frac{i\omega_0}{2 \epsilon_0} \mpol_q ,
\label{eq_rt0}
\end{equation}
where $\mpol_q$ is the complex amplitude of the macroscopic polarization that
relates to $\tilde{P}_q$ as $\td_q$ relates to $\tilde{E}_q$. The form of eq.(\ref{eq_rt0}) is
in accord with \citet{1973ApJ...179..111G}, who also include a macroscopic magnetization.

\subsubsection{Macroscopic Polarization}
\label{sss:macropol}

The macroscopic polarization is the velocity-integrated expectation value of the electric dipole
operator of the active molecule that amplifies the maser. It is a reasonably
general result from quantum-mechanics that such an expectation value is the
trace of the matrix product of the molecular density matrix (DM) and the operator.
Writing the dipole operator as $\hat{\mtx{d}}$, the expectation value
is therefore $\langle \hat{\mtx{d}} \rangle = \mathrm{Tr} [\uprho \hat{ \mtx{d} }]$,
where $\uprho (\vec{r},t,\vec{v})$ is the molecular DM, a function
of molecular velocity $\vec{v}$ as well as position and time. Note that the
row and column indices of these matrices correspond to energy levels of the
molecule and that, in the case of the dipole, its individual elements,
$\hv{d}_{pq}$ are themselves vectors. To obtain the macroscopic polarization we
must integrate over the molecular velocity to obtain,
\begin{equation}
\vec{P}(\vec{r},t) = \int_{\vec{v}} d^3v \mathrm{Tr}[\uprho (\vec{r},t,\vec{v})\hat{\mtx{d}}] .
\label{eq_macropol}
\end{equation}

By isolating one component of the macroscopic polarization, changing to an analytic
signal representation, and using the standard representation of the trace of a
matrix product in terms of individual elements, one Cartesian component of the
complex amplitude of $\vec{P}$ may be written as
\begin{equation}
\mpol_q = 2 e^{iY_0} \sum_{p=2}^{N} \sum_{k=1}^{p-1} \hat{d}_{pk,q}^* \int_{\vec{v}} \rho_{pk} d^3v.
\label{eq_macropol2}
\end{equation}
where $p$ and $q$ represent molecular energy levels from a total of $N$, and
$\hat{d}_{pk,q}$ is the Cartesian component $q$ of the element $pq$ of the density
matrix. To remove the rapidly varying factor in eq.(\ref{eq_macropol2}) we
write the off-diagonal element of the DM as the product of a slowly varying
part $s_{pk}$ and a rapidly oscillating term as follows \citep{1978PhRvA..17..701M}:
\begin{equation}
\rho_{pk}(\vec{r},t,v) = -(i/2) s_{pk}(\vec{r},t,v) e^{-iY_0} .
\label{eq_sdefn}
\end{equation}
Substitution of eq.(\ref{eq_sdefn}) into eq.(\ref{eq_macropol2}), and substiution of the
result in turn into eq.(\ref{eq_rt0}) yields the radiative transfer equation,
\begin{equation}
\left( \frac{\pd}{\pd t} + c \hv{n} \cdot \del \right) \td_q(\vec{r},t)
=
\frac{\omega_0}{2\epsilon_0} \sum_{p=2}^{N} \sum_{k=1}^{p-1} \hat{d}_{pk,q}^* \int_{\vec{v}} s_{pk}(\vec{r},t,\vec{v}) d^3v .
\label{eq_rt2}
\end{equation}

\subsubsection{Simplifications}
\label{sss:simp}

So far, we have kept our description of the electric field and its transfer
very general. At this point, however, we introduce some useful simplifications
resulting from the geometry adopted in Section~\ref{axes}. The radiation is assumed
to propagate along the $z$ axis, in the direction of increasing $z$: therefore
$\hv{n} = \hv{z}$. We can therefore reduce our transfer equation, eq.(\ref{eq_rt2}) to
\begin{equation}
d_t \td_q(r,\phi,z,t)
=
\frac{\omega_0}{2\epsilon_0} \sum_{p=2}^{N} \sum_{k=1}^{p-1} \hat{d}_{pk,q}^* \int_{-\infty}^\infty s_{pk}(r,\phi,z,t,v) dv .
\label{eq_rt3}
\end{equation}
The derivative on the left-hand side of eq.(\ref{eq_rt3}) is the shorthand notation,
\begin{equation}
d_t = \pd /\pd t + c \pd/\pd z ,
\label{eq_shortdt}
\end{equation}
the general position $\vec{r}$ has been replaced by the global 
$(r,\phi,z)$ cylindrical coordinates
of Section~\ref{axes}, and we now consider only the $z$-component of the
molecular velocity, $v=v_z$, since only the Doppler effect along the
axis of propagation is observable. It is straightforward to show
that eq.(\ref{eq_rt3}) applies to spherical, as well as Cartesian, components.

In the definition of the electric field, we re-define $Y_0$ as
\begin{equation}
Y_0 = \omega_0 ( t - z/c) ,
\label{eq_newbigY}
\end{equation}
and with this modification, we may still use eq.(\ref{eq_sig1}) and eq.(\ref{eq_sig2})
to represent the analytic signal. We will also sometimes need a form in
Cartesian components, but in terms of the spherical complex amplitudes, for example
\begin{equation}
\tv{E}(r,\phi,z,t) = 2^{-1/2}[
        \hv{x}(\td_R + \td_L) + i \hv{y} (\td_R - \td_L) + \sqrt{2}\hv{z} \td_z
                     ] e^{-iY_0} .
\label{eq_cartsig}
\end{equation}

\subsection{Molecular Response}
\label{response}
Equations for the evolution of general diagonal and off-diagonal elements
of the molecular DM are taken from \citet{mybook}, where
they are derived in detail from Schr\"{o}dinger's equation. For the
diagonal element, $\rho_{qq}$,
\begin{equation}
D_t\rho_{qq} = \frac{i}{\hbar} \sum_{j=1}^N (\rho_{qj} \ham_{jq} - \rho_{jq} \ham_{qj})
                            +\sum_{j=1}^N (k_{jq} \rho_{jj} - k_{qj} \rho_{qq}),
\label{eq_diag0}
\end{equation}
noting that a diagonal element represents the number density of molecules
in level $q$, or, in a normalised form, the probability of occupancy of
level $q$. The equation also makes use of off-diagonal elements of the
DM, for example $\rho_{pq}$, where $p\neq q$, that represent 
coherence between pairs of levels from the total of $N$. Coupling to the maser
radiation field is delivered through the matrix elements of the interaction
Hamiltonian, $\ham_{pq}$, and to other forms of level-changing process, such
as kinetic collisions, via the all-process rate coefficients, $k_{pq}$.
The total derivative is now,
\begin{equation}
D_t = \pd/\pd t + v \pd /\pd z  ,
\label{eq_vderiv}
\end{equation}
noting that a molecular $z$-velocity, $v$, now replaces the speed of light
used in radiative transfer equations (see eq.(\ref{eq_shortdt})). The diagonal
elements of the DM have the functional dependence,
\begin{equation}
\rho_{qq} = \rho_{qq}(r,\phi,z,t,v) .
\label{eq_rhofuncdep}
\end{equation}

The general off-diagonal element, $\rho_{pq}$, evolves according to the equation
\begin{equation}
D_t\rho_{pq}  = \frac{i}{\hbar} \sum_{j=1}^N (\rho_{pj} \ham_{qj}^* - \rho_{jq} \ham_{pj})
                            - i \omega_{pq} \rho_{pq} - \rho_{pq} / \tau_{pq} .
\label{eq_off0}
\end{equation}
A complex-conjugate version of an interaction Hamiltonian element has been
used, noting that, like the DM, $\hat{\mathrm{H}}$ is Hermitian. In 
eq.(\ref{eq_off0}) we also take $p$ ($q$) to be the upper (lower) level
of the pair, so that the angular frequency $\omega_{pq}$, corresponding to
the transition energy between the levels, is positive, and
$\omega_{qp} = -\omega_{pq}$. The timescale $\tau_{pq}$ is the timescale over
which coherence in the transition $pq$ is lost. This will in general be
shorter than $1/k_{pq}$ because it includes elastic processes, such as
collisions that change molecular direction but not level. The functional
dependence of the off-diagonal element is the same as for the diagonal
element in eq.(\ref{eq_rhofuncdep})

\subsection{Off-Diagonal Equation: Modifications}
\label{ss:odm}
We define the population inversion between upper
level $p$ and lower level $q$ as
\begin{equation}
\Delta_{pq}(r,\phi,z,t,v) = \rho_{pp} - \rho_{qq} ,
\label{eq_invdef}
\end{equation}
and isolate it from off-diagonal elements.
To do this, we extract from the sum in eq.(\ref{eq_off0}) those terms where
$j=p$ and $j=q$, and write them separately. Since elements of
the interaction Hamiltonian are defined by the equation,
\begin{equation}
\ham_{pq}(r,\phi,z,t) = - \vec{E}(r,\phi,z,t) \cdot \hv{d}_{pq} ,
\label{eq_inthamdef}
\end{equation}
any element of the form $\ham_{jj}=0$ because the dipole $\hv{d}_{jj}=0$. Using this fact, and with the help of
eq.(\ref{eq_vderiv}), eq.(\ref{eq_invdef}) and the Hermitian property of $\ham_{qp}$,
we may write eq.(\ref{eq_off0}) in the modified form,
\begin{align}
D_t\rho_{pq} &= 
\frac{i}{\hbar} \sum_{j\neq p,q}^N (\rho_{pj} \ham_{qj}^* - \rho_{jq} \ham_{pj}) \nonumber \\
            &+ \frac{i\ham_{pq}\Delta_{pq}}{\hbar}
                            - (i \omega_{pq}  + \gamma_{pq}) \rho_{pq} ,
\label{eq_off2}
\end{align}
where the homogeneous line width $\gamma_{pq} = 1/T_{pq}$.

\subsection{Diagonal Equation: Modifications}
\label{ss:dem}
To replace equations describing the evolution
of individual level populations with equations that describe the evolution
of an inversion, we write down a version of eq.(\ref{eq_diag0}) in
which $q$ is replaced by $p$, and subtract the original eq.(\ref{eq_diag0}) from
it. Equation~\ref{eq_invdef} dictates that the left-hand side of the result becomes
the differential of the inversion $\Delta_{pq}$. Using the Hermitian property of both
the interaction Hamiltonian and DM, the result of the subtraction is
\begin{align}
D_t \Delta_{pq} & = \frac{-2}{\hbar} \sum_{j=1}^N ( \Im \{ \rho_{pj} \ham_{jp} \} - \Im \{ \rho_{qj} \ham_{jq} \} )
               + \sum_{j=1}^N \rho_{jj} \Delta k_{j,pq} \nonumber \\
               & - \rho_{pp} k_{p\Sigma} + \rho_{qq} k_{q\Sigma},
\label{eq_diag2}
\end{align}
where we have defined $k_{p\Sigma} = \sum_{j=1}^N k_{pj}$ for the total
rate-coefficient out of level $p$ (and similarly for level $q$), and
where $\Delta k_{j,pq} = k_{jp} - k_{jq}$. On the basis that the total rate coefficient out
of any level is approximately the same, we make the approximation $k_{p\Sigma} \sim k_{q\Sigma} \sim \Gamma_{pq}$,
where $\Gamma_{pq}$ is the transition loss rate. The term in $\Delta k_{j,pq}$ represents a source term for
the inversion, arising from all the remaining levels $\rho_{jj}$. We set it equal to a 
phenomenological pumping term,
\begin{equation}
\sum_{j=1}^N \rho_{jj} \Delta k_{j,pq} = P_{pq} \phi (\vec{v}) ,
\label{eq_phenpump}
\end{equation}
where $P_{pq}$ is the pumping constant for the $pq$ inversion and $\phi (\vec{v})$ is the
molecular velocity distribution function. The appearance of this distribution function in
eq.(\ref{eq_phenpump}) is justified on the grounds that the left-hand side
contains the populations $\rho_{jj}(r,\phi,z,t,\vec{v})$ that all follow this distribution.

Extracting the $j=p$ and $j=q$ terms from the sum in eq.(\ref{eq_diag2}), and 
employing the Hermitian property of $\hat{\mtx{H}}$, together with
eq.(\ref{eq_vderiv}), eq.(\ref{eq_invdef}) and eq.(\ref{eq_phenpump}), the inversion
evolves according to
\begin{align}
D_t \Delta_{pq} &=
-\frac{2}{\hbar} \Im \left\{
                             2 \rho_{pq} \ham_{qp} 
                             + \sum_{j \neq p,q}^N (\rho_{pj} \ham_{jp} - \rho_{qj} \ham_{jq})
                     \right\} \nonumber \\
              & + P_{pq} \phi (\vec{v}) - \Gamma_{pq} \Delta_{pq} .
\label{eq_diag3}
\end{align}

From here on, we will assume that the molecular velocity distribution is a Gaussian
function of the form,
\begin{equation}
\phi (\vec{v}) = (\pi^{1/2} w)^{-1} e^{-v^2/w^2 ,}
\label{eq_gaussian}
\end{equation}
where $w$ is a width parameter given by
\begin{equation}
w = (2k_B T_K /m_X)^{1/2} ,
\label{eq_gwidth}
\end{equation}
for kinetic temperature $T_K$ and molecular mass of the maser molecule, $m_X$.
Note that $\phi(\vec{v})$ can be changed only by molecular collisions and not by
the Zeeman effect or by saturation or any other process within the maser.

\subsection{Zeeman Patterns}
\label{ss:zeeman}
We consider a weak-field Zeeman effect in which the
the energy shift is proportional to the external field strength.
Much more complicated patterns have been considered, for example by
\citet{2006ApJ...636..548A}, but not in connection with masers.
We will also assume that the (local) magnetic field direction provides
a good quantization axis. We will attempt to use a generic system
where the maser molecule has a magnetic moment
\begin{equation}
\vec{m}_J = - \mu_X g_J \vec{J},
\label{eq_magmoL}
\end{equation}
where the magneton, $\mu_X$, is the nuclear magneton, $e\hbar/(2m_p)$,
for closed-shell molecules. For molecules like OH, with net electronic
angular momentum, the Bohr magneton is used, replacing the proton
mass $m_p$ with the electron mass $m_e$ in the formula above. The
angular momentum $\vec{J}$ would be replaced by $\vec{F}$ in the
case of OH \citep{1955PhRv..100.1735D}. We also assume that the
Land\'{e} factor, $g_J$, is a positive number in eq.(\ref{eq_magmoL}),
so that the magnetic moment lies anti-parallel to $\vec{J}$. Our Zeeman
pattern therefore follows, in sign, the usual `textbook' case of
electronic angular momentum. If $\vec{J}$ results mostly from molecular
rotation, we are therefore assuming that the contribution to $g_J$ from
the coupling to the electron cloud has a greater magnitude than
the contribution from the nuclear framework.

The Zeeman Hamiltonian for a molecule with magnetic moment $\vec{m}_J$ in an
external magnetic field $\vec{B}$ is
\begin{equation}
H_Z = - \vec{m} \cdot \vec{B}
\label{eq_zeeham}
\end{equation}
\citep{1980woodgate.book.L,eisberg1985quantum,mybook}. The resulting Zeeman
energy shift is
\begin{equation}
\Delta E (J,M) = \mu_X g_J B M ,
\label{eq_zeeshift}
\end{equation}
where $B(x,y,z)=|\vec{B}|$ is the local magnetic field strength, and
$M$ is the magnetic quantum number that has integer values
in the range $-J..0..+J$, where $J$ is the quantum number corresponding 
to $\vec{J}$. An unsplit level denoted by $J$ is therefore split by the
external magnetic field into $2J+1$ equally spaced Zeeman sublevels.

\subsubsection{Transition Types}
\label{sss:ztrans}
\begin{figure}
\includegraphics[width=84mm,angle=0]{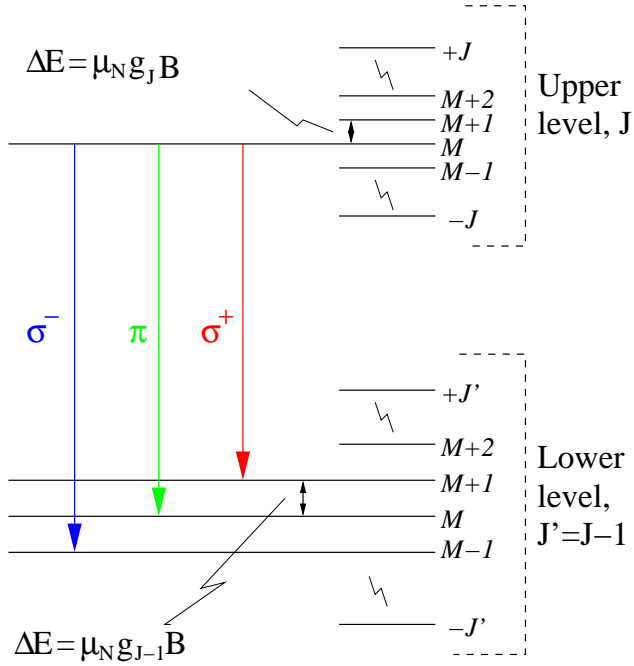}
\caption{The Zeeman sub-levels of the two-level system considered in
this work; the magnetic moment is anti-parallel to $\vec{J}$.
The three possible electric-dipole transitions in emission
from sublevel $M$ of the upper rotational state are shown as coloured
vertical arrows. Note that there is only one such
transition of each type. Zeeman energy shifts are also shown.}
\label{fig_trantypes}
\end{figure}

We will consider a pair of unsplit levels:
an upper level $J$ and lower level $J'=J-1$, and take an idealised
case where these are isolated from any other levels. Both of these unsplit
levels then contain
magnetic sublevels $M$, ranging from $-J$ to $+J$ in the upper level
and from $-J'$ to $+J'$ in the lower level.
Transitions between
magnetic sub-levels of $J$ and $J'$ are subject to the electric dipole
selection rules: $\Delta M = 0,\pm1$ (Figure~\ref{fig_trantypes}).

The allowed transitions will be classified as follows: let $\Delta M$
be the change in $M$ in emission. A transition with $\Delta M =0$ will
be termed a $\pi$-transition. Transitions with $\Delta M = +1$
($\Delta M = -1$) will be called $\sigma^+$ ($\sigma^-$) transitions.
Note that any sublevel $M$ in $J$ can have at most one 
electric-dipole allowed transition of
each type leaving it for some destination magnetic sub-level within $J'$ 
(see Figure~\ref{fig_trantypes}). Note that this classification of the $\sigma$
transitions is not universal and that 
\citet{1973ApJ...179..111G}, in particular, use 
the reverse definition.

The energy of our upper energy level with quantum numbers $J,M$, relative to
a gound level of zero, is just the energy $E_J$ of the unsplit rotational level
added to the Zeeman shift from eq.(\ref{eq_zeeshift}):
\begin{equation}
E(J,M) = E_J + \mu_X g_J B M ,
\label{eq_Eupper}
\end{equation}
whilst the energies of the three target sub-levels in $J'=J-1$ become
\begin{subequations} \label{eq_Elower}
\begin{align}
E(J-1,M) = E_{J-1} + \mu_X g_{J-1} B M \label{eq_Elower_pi}\\
E(J-1,M\pm1) = E_{J-1} + \mu_X g_{J-1} B (M\pm1). \label{eq_Elower_sigma}
\end{align} 
\end{subequations}
The frequency of the $\pi$-transition, $\nu_M^0$, may be found by subtracting 
eq.(\ref{eq_Elower_pi}) from eq.(\ref{eq_Eupper}) and dividing through
by Planck's constant. If we let the frequency of the unsplit transition
be $\nu_0 = (E_J - E_{J-1})/h$ then we find
\begin{equation}
\nu_M^0 = \nu_0 + \mu_X B M (g_J - g_{J-1})/h ,
\label{eq_pifreq}
\end{equation}
noting that, for a $\pi$-transition, the frequency reverts to that of the
unsplit transition if either $M=0$ (the `central $\pi$-transition') or
if the Land\'{e} factors of the two rotational levels are equal. In this
latter case, the frequencies of all the $\pi$-transitions are the same,
resulting in a single $\pi$-spectral line centered on $\nu_0$. The
frequencies of the $\sigma$-transitions may be constructed in a similar
manner, and the equation analogous to eq.(\ref{eq_pifreq}), summarizing
both types, is
\begin{equation}
\nu_M^\pm = \nu_0 + \mu_X B [M(g_J - g_{J-1}) \mp g_{J-1}] /h ,
\label{eq_sigfreq}
\end{equation}
where the upper (lower) optional signs refer to $\sigma^+$ ($\sigma^-$) transitions. Note
that a single frequency for each type again results, regardless of $M$, if
both Land\'{e} factors are equal.

\subsubsection{Dipole Orientation}
\label{sss:dipor}
\begin{figure}
\includegraphics[width=84mm,angle=0]{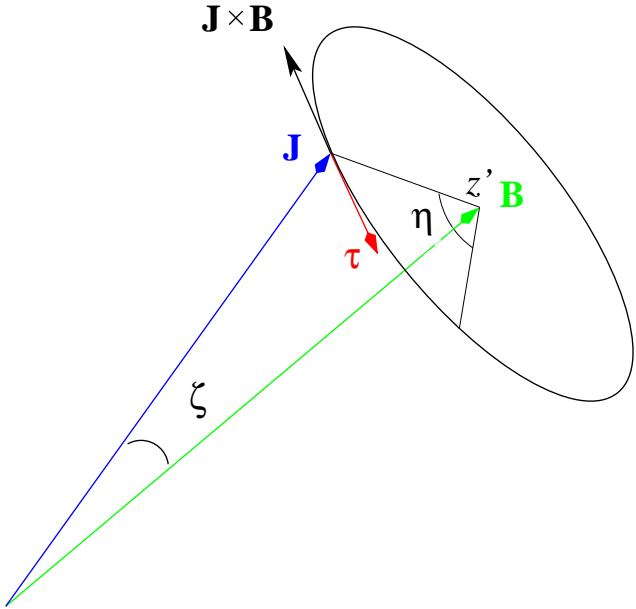}
\caption{The torque on the molecular angular momentum vector due to
a magnetic field along the $z'$ axis: the vector cross product
$\vec{J} \vprod \vec{B}$ points tangentially in the direction shown
by definition. The magnitude of $\vec{J}$ and the angle $\zeta$ it
makes to the magnetic field must remain constant in the absence of
transition, so the head of the vector $\vec{J}$ moves in a circle, shown
in projection as an ellipse, around the magnetic field. The torque
exerted on $\vec{J}$, shown in red, is opposite in direction to the
cross product from eq.(\ref{eq_torque}). As the torque would still
be tangential after advancing the head of $\vec{J}$ through an angle
$\eta$, the motion of $\vec{J}$ is generally anticlockwise as
viewed by an observer `receiving' the magnetic field, and therefore
described by the spherical vector $\hv{e}_R'$ under the IEEE convention.}
\label{fig_torque}
\end{figure}

The aim of this section is to relate the helical type of a 
magnetically-split transition to a particular spherical basis vector
in the primed (magnetic field based) frame. The external magnetic
field exerts a torque on the magnetic moment, equal to
\begin{equation}
\vec{\tau} = \vec{m} \vprod \vec{B} = - \mu_X g_J \vec{J} \vprod \vec{B}
\label{eq_torque}
\end{equation}
\citep{1972engewehrrichards}. The second form on the right-hand side of
eq.(\ref{eq_torque}) results from elimination of the magnetic moment with
the aid of eq.(\ref{eq_magmoL}). With the aid of a diagram, Fig.~\ref{fig_torque},
we can see that the torque is generally anticlockwise from the point of
view of an IAU receiver-based observer (see Fig.~\ref{fig_iau_axes}), and therefore right-handed
in the IEEE convention. As we are in the magnetic field (primed) frame, we
can now say that the torque, and therefore the precessional
motion of $\vec{J}$, is proportional to the right-handed
spherical unit vector $\hv{e}_R'$. A supporting analysis is given in
\citet{1979anpa.book.....L}.

Having established that the direction of the precession of $\vec{J}$ about
$\vec{B}$ (or the $z'$ axis) is IEEE-right-handed, it is time to introduce the effect of transitions. For
the sake of example, we shall consider a $\sigma^+$ transition, in fact the
one marked by the red emission arrow in Fig.~\ref{fig_trantypes}.
$M$ is the quantum number for the projection of $\vec{J}$ on the z' axis,
and this has increased by one unit in transition: that is, a $\sigma^+$ transition
in emission adds one unit of IEEE right-handed angular momentum about the
$z'$ axis to the molecule. Since the transition is mediated by the electric
dipole of the molecule, the dipole matrix element for a $\sigma^+$ transition
in emission is also IEEE right-handed. This discussion is summarized and
extended to stimulated emission in \citet{2014MNRAS.440.2988G}.
Angular momentum about $z'$ is conserved by the emission of an IEEE left-handed
photon in the direction of increasing $z'$, that is towards the observer.
This conclusion is in agreement with the classical Lorentzian results, backed
up by over 100\,years of laboratory experiments, going back to the
days of Zeeman himself. We have
a magnetic field pointing towards the observer (Fig.~\ref{fig_torque}), and
the $\sigma$ transition with the lower frequency (the $\sigma^+$) will be
observed as emitting IEEE left-hand circularly polarized radiation (or  the
$\sigma$ transition with the higher frequency ($\sigma^-$) will be observed
emitting IEEE right-hand circularly polarized radiation. 

In Section~\ref{sss:macropol},
we introduced the dipole matrix element for a transition between
upper level $p$ and lower level $q$ as
$\hv{d}_{pq}$. We now need to replace this very general notation
with something more applicable to the Zeeman group structure
depicted in Figure~\ref{fig_trantypes}. Our upper level can be
identified by the quantum numbers $(J,M)$, while the three possible
options for the lower level are $(J',M+1)$ for $\sigma^+$,
$(J',M)$ for $\pi$ or $(J',M-1)$ for $\sigma^-$ transitions. Now we
have shown above that, for the $\sigma^+$ transition type, the
dipole is right-handed in the magnetic-field based (primed) axis system.
For example,
\begin{equation}
 \hv{d}_{pq} = \hv{d}_{(J,M),(J',M+1)} = \hat{d}_{(J,M),(J',M+1)} \hv{e}_R' ,
\label{eq_dipex}
\end{equation}
for a $\sigma^+$ dipole in the Zeeman group notation.
By extension, the dipole of a $\sigma^-$ transition is IEEE left-handed.
It is well known that for radiation propagation parallel to the
magnetic field, the $\pi$-transition does not appear. This means that
the dipole has no component in the $x'y'$-plane, where it could interact
with the electric field of the radiation, and must therefore lie
along the $z'$ axis. We may now write down three dipole definitions:
\begin{subequations} \label{eq_dipgold}
\begin{align}
\sigma^+ &: \hv{d}_{pq} = \hv{d}_{(J,M),(J',M+1)} = \hv{d}_M^+ = \hat{d}_M^+ \hv{e}_R' \label{eq_dipgold_sp}\\
\pi      &: \hv{d}_{pq} = \hv{d}_{(J,M),(J',M)} = \hv{d}_M^0 = \hat{d}_M^0 \hv{z}' \label{eq_dipgold_pi}\\
\sigma^- &: \hv{d}_{pq} = \hv{d}_{(J,M),(J',M-1)} = \hv{d}_M^- = \hat{d}_M^- \hv{e}_L' ,\label{eq_dipgold_sm}
\end{align}
\end{subequations}
where the final forms introduce a useful shorthand that avoids the use
of an excessive number of subscripts. It is unambiguous because: (i) all
transitions are from $J$ to $J'$, and (ii) there is only one
transition of each type leaving sublevel $M$ in $J$. In the shorthand,
the only subscript, $M$, still denotes the magnetic quantum number of
the upper level, whilst the superscript indicates the change in $M$
in emission.

\subsubsection{Dipoles in the Radiation Frame}
\label{sss:diprad}
Having defined the dipoles as pure helical components in the primed
frame (equation~\ref{eq_dipgold}), we also derive a representation of
the dipoles in the radiation-based (global, unprimed) frame. This is straightforwardly
achieved by applying the rotation matrix, eq.(\ref{eq_sphgold}), to the primed unit
vectors in eq.(\ref{eq_dipgold}), yielding
\begin{subequations} \label{eq_diprad}
\begin{align}
\hv{d}_M^+ &= \hat{d}_M^+ [ \hv{e}_R (1+\cos \theta)(c-is)/2 \nonumber \\
          &+\hv{z} (i/\sqrt{2}) \sin \theta + \hv{e}_L (1-\cos \theta)(c+is)/2] \\
\hv{d}_M^0 &= (\hat{d}_M^0\sqrt{2}) [ \hv{e}_R (s+ic)\sin \theta \nonumber \\
          &+ \sqrt{2}\hv{z} \cos \theta + \hv{e}_L (s-ic)\sin \theta ] \\
\hv{d}_M^- &= \hat{d}_M^- [ \hv{e}_R (1-\cos \theta)(c-is)/2 \nonumber \\
          &- \hv{z} (i/\sqrt{2}) \sin \theta + \hv{e}_L (1+\cos \theta)(c+is)/2] ,
\end{align}
\end{subequations}
where we have used the shorthand notation $c+is = \cos \phi' + i \sin \phi'$, and similar
bracketed expressions. We also convert the dipole elements in eq.(\ref{eq_diprad})
to the Cartesian basis, using the definitions
in eq.(\ref{eq_rldef_R}) and eq.(\ref{eq_rldef_L}) for the conversion:
\begin{subequations} \label{eq_dipcart}
\begin{align}
\hv{d}_M^+ &= (\hat{d}_M^+/\sqrt{2}) [ \hv{x} (\cos \phi' - i\cos \theta \sin \phi') \nonumber \\
          &+ \hv{y} (\sin \phi' + i\cos \theta \cos \phi') + i \hv{z} \sin \theta] \label{eq_dipcart_sp}\\
\hv{d}_M^0 &= \hat{d}_M^0 [ \hv{x} \sin \theta \sin \phi' -\hv{y} \sin \theta \cos \phi' + \hv{z} \cos \theta ] \label{eq_dipcart_pi} \\
\hv{d}_M^- &= (\hat{d}_M^-/\sqrt{2}) [ \hv{x} (\cos \phi' + i\cos \theta \sin \phi') \nonumber \\
          &+ \hv{y} (\sin \phi' -i \cos \theta \cos \phi') - i \hv{z} \sin \theta] .\label{eq_dipcart_sm}
\end{align}
\end{subequations}

\section{Time-Domain Equations}
\label{tdom}

\subsection{Interaction Hamiltonian}
\label{ss:hint}
Elements of the interaction Hamiltonian have been previously introduced using
the general transition $pq$ in eq.(\ref{eq_inthamdef}), so the indices simply follow those of the
matrix element of the dipole. We can therefore adopt the same shorthand notation
for the elements relevant to our Zeeman group: $H_M^\pm$ for the $\sigma^\pm$ transitions
and $H_M^0$ for the $\pi$-transitions. In the $\sigma^+$ transition out of sub-level $M$, for example,
the Hamiltonian element is $H_M^+ = -\vec{E} \cdot \hv{d}_M^+$. For the electric field, we
take the real part of the Cartesian broad-band analytic signal, eq.(\ref{eq_cartsig}).
The results of the dot product with the three forms of eq.(\ref{eq_dipcart}) are,
\begin{subequations} \label{eq_hamgold}
\begin{align}
H_M^+ &= \{ [(\td_R + \td_L) e^{-iY_0} + c.c.] (\cos \phi' - i \sin \phi' \cos \theta ) \nonumber \\
       &+ [i(\td_R - \td_L) e^{-iY_0} + c.c.] ( \sin \phi' + i \cos \phi' \cos \theta \nonumber \\
       &+\sqrt{2}i[\td_z e^{-iY_0} + c.c.] \sin \theta \} (-d_M^+/4) \label{eq_hamgold_sp}\\
H_M^0 &= \{[[(\td_R + \td_L) e^{-iY_0} + c.c.] \sin \phi' \sin \theta \nonumber \\
       &-[i(\td_R - \td_L) e^{-iY_0} + c.c. ]\cos \phi' \sin \theta \nonumber \\
       &+\sqrt{2}[\td_z e^{-iY_0} + c.c.] \cos \theta \} (-\sqrt{2}d_M^0/4)\label{eq_hamgold_pi} \\
H_M^- &= \{ [(\td_R + \td_L) e^{-iY_0} + c.c.] (\cos \phi' + i \sin \phi' \cos \theta ) \nonumber \\
       &+[i(\td_R - \td_L) e^{-iY_0} + c.c.] (\sin \phi' -i \cos \phi' \cos \theta )  \nonumber \\
       &-\sqrt{2}i[\td_z e^{-iY_0} + c.c.] \sin \theta \} (-d_M^-/4)\label{eq_hamgold_sm}
\end{align}
\end{subequations}
where $c.c.$ denotes the complex conjugate.

\subsection{Off-Diagonal Equations}
\label{ss:offtime}
The Hamiltonian matrix elements in eq.(\ref{eq_hamgold}) are specific to magnetic transition
types, so we must develop sets of equations for the off-diagonal and diagonal DM
elements that are similarly type-specific. Beginning with the off-diagonal DM
elements that evolve according to eq.(\ref{eq_off2}), we consider membership of the sum
over $j$. We will assume that off-diagonal elements of the DM are negligible
unless they correspond to an allowed electric dipole transition. This assumption is
frequently, but not universally, made. Such elements are set to zero in
\citet{2009MNRAS.399.1495D}, but \citet{1973ApJ...179..111G} make them constants in their analysis, whilst
discussing the possibility that they may be comparable to those elements that correspond
to the dipole transitions. 

If, as we shall assume, the off-diagonal elements of the DM that correspond
to forbidden electric dipole transitions are set to zero, an important simplification
results: the sum over $j$ in eq.(\ref{eq_off2}) is empty for the Zeeman group introduced
in Section~\ref{sss:ztrans} \citep{mybook}. The result is that off-diagonal 
DM elements for all three transitions out of level $(J,M)$ to magnetic
sublevels in $J'$ all evolve according to the generic equation,
\begin{equation}
D_t\rho_M = i\ham_M \Delta_M /\hbar - (\gamma_M + i \omega_M ) \rho_M 
\label{eq_off3}
\end{equation}
The off-diagonal elements of the DM have already been expanded as the
product of a fast and a slow term in eq.(\ref{eq_sdefn}), and we use this
equation to eliminate $\rho_M$ from eq.(\ref{eq_off3}) in favour of the
slowly-varying $s_M$. The result is
\begin{equation}
D_t s_M = -2\ham_M \Delta_M e^{iY_0}/\hbar - [\gamma_M - i (\omega_0 - \omega_M -v\omega_0/c)] s_M .
\label{eq_off4}
\end{equation}
We complete the derivation of type-dependent equations for the evolution of
$s_M^\pm, s_M^0$ by introducing the appropriate forms of eq.(\ref{eq_hamgold})
into eq.(\ref{eq_off4}), and applying the rotating wave equation to remove
terms oscillating rapidly at $e^{\pm i Y_0}$. The evolution equations are,
\begin{subequations} \label{eq_offtime}
\begin{align}
D_t s_M^\pm &= \frac{\Delta_M^\pm\hat{d}_M^\pm}{2\hbar} [
    (\td_R+\td_L) (c' \mp is'\cos \theta) \nonumber \\
         &+ i(\td_R-\td_L)(s' \pm i c' \cos \theta) \pm \sqrt{2}i\td_z \sin \theta
                                               ] \nonumber \\
         &- [\gamma_M^\pm +i (\Delta \omega_M^\pm + v\omega_0/c) ] s_M^\pm \label{eq_offtime_sig} \\
D_t s_M^0 &= \frac{\Delta_M^0\hat{d}_M^0}{\sqrt{2}\hbar} [
    (\td_R+\td_L) s' \sin \theta \nonumber \\
         &- i(\td_R-\td_L) c' \sin \theta + \sqrt{2}\td_z \cos \theta
                                               ] \nonumber \\
         &- [\gamma_M^0 +i (\Delta \omega_M^0 + v\omega_0/c) ] s_M^0 \label{eq_offtime_pi} 
\end{align}
\end{subequations}
where $c'=\cos \phi'$, $s' = \sin \phi'$, and the equations for both $\sigma$-transitions have
been combined into the single equation, eq.(\ref{eq_offtime_sig}) with optional signs:
the upper (lower) version of such signs refers to $\sigma^+$ ($\sigma^-$).
Zeeman shifts are defined as
\begin{subequations} \label{eq_zshift}
\begin{align}
\Delta \omega_M^\pm &= \mu_X B [M(g_J - g_{J−1}) \mp g_{J−1}]/\hbar \label{eq_zshift_sig} \\
\Delta \omega_M^0 &= \mu_X BM (g_J - g_{J−1})/\hbar , \label{eq_zshift_pi}
\end{align}
\end{subequations}
with the help of eq.(\ref{eq_sigfreq}) and eq.(\ref{eq_pifreq}), respectively.

\subsection{Inversion Equation}
\label{ss:diagtime}

The sum in the equation for the evolution of the inversion, eq.(\ref{eq_diag3}), is
not empty when we convert to the Zeeman group notation used in Section~\ref{ss:offtime}.
Instead, they introduce interaction contributions from neighbouring transitions of
all three polarization types, including those different from the type of the
inversion in the differential. Membership of the sums is established in \citet{mybook} and
results in the following type-specific developments of eq.(\ref{eq_diag3}):
\begin{subequations} \label{eq_diag4}
\begin{align}
 D_t \Delta_M^\pm &= P_M^\pm \phi(v) - \Gamma_M^\pm \Delta_M^\pm -\frac{2}{\hbar} \Im \left\{
    2\rho_M^\pm \ham_M^{\pm *} + \rho_M^0 \ham_M^{0*} \right. \nonumber \\
   &+ \left. \rho_M^\mp \ham_M^{\mp *} - \rho_{M+1}^{0*} \ham_{M+1}^0 - \rho_{M+2}^{\mp *} \ham_{M+2}^\mp
                        \right\} \label{eq_diag4_sig}\\
D_t \Delta_M^0 &= P_M^0 \phi(v) - \Gamma_M^0 \Delta_M^0 -\frac{2}{\hbar} \Im \left\{
   2\rho_M^0 \ham_M^{0*} + \rho_M^+ \ham_M^{+*} \right. \nonumber \\
   &+ \left. \rho_M^- \ham_M^{-*} - \rho_{M-1}^{+*} \ham_{M-1}^+ - \rho_{M+1}^{-*} \ham_{M+1}^-
                        \right\} ,\label{eq_diag4_pi}
\end{align}
\end{subequations}
where eq.(\ref{eq_diag4_sig}) refers to $\sigma$-transitions. The upper (lower) optional
signs apply to $\sigma^+$ ($\sigma^-$). The case of $\pi$-transitions is covered by
eq.(\ref{eq_diag4_pi}).

All of the interaction terms in eq.(\ref{eq_diag4}) may be written in the general
form $\rho_x^a \ham_x^{a*}$, or its complex conjugate,  where $M-2\leq x \leq M+2$ is an upper magnetic sublevel, and
$a=0,\pm$ is a transition type. The general interaction term may be developed
by eliminating the Hamiltonian elements in favour of expressions from eq.(\ref{eq_hamgold}),
or complex conjugates thereof, and by using eq.(\ref{eq_sdefn}), its conjugate, and the
rotating wave approximation to remove all rapidly varying parts of the DM elements.
We eventually obtain, for specific values of $a$,
\begin{subequations} \label{eq_itype1}
\begin{align}
\rho_x^\pm \ham_x^{\pm*} &= i\hat{d}_x^{\pm*} s_x^\pm [
  (\td_R^* + \td_L^*)(c' \pm is' \cos \theta ) \nonumber \\ 
&-i(\td_R^* - \td_L^*)(s' \mp ic' \cos \theta ) \mp \sqrt{2}i\td_z^* \sin \theta 
                                                ]/8 \label{eq_itype1_sig}\\
\rho_x^0 \ham_x^{0*} &= \sqrt{2} i\hat{d}_x^{0*} s_x^0 [
    (\td_R^* + \td_L^*) s' \sin \theta  \nonumber \\
& + i(\td_R^* -\td_L^*) c' \sin \theta + \sqrt{2} \td_z^* \cos \theta
                                                   ]/8 . \label{eq_itype_pi}
\end{align}
\end{subequations}
With appropriate choices for $x$, the various forms of eq.(\ref{eq_itype1}) may be
used to eliminate the interaction terms from eq.(\ref{eq_diag4}). The resulting,
rather cumbersome, expressions are the evolution equations for the inversion
in the three different types of transition. For brevity, we again combine the
expressions for $\sigma$ transitions into a single equation, with the upper form
of any optional sign referring to $\sigma^+$. Inversions evolve according to
\begin{subequations} \label{eq_invtime}
\begin{align}
\!\!\!& D_t \Delta_M^\pm  = P_M^\pm \phi(v) - \Gamma_M^\pm \Delta_M^\pm -\frac{1}{4\hbar} \Im \{ \nonumber \\
   &  \sqrt{2} i s' \!\sin \theta [\hat{d}_M^{0*} s_M^0 (\td_R^* \!+\! \td_L^*) + \hat{d}_{M\!\pm \!1}^{0} s_{M \!\pm \!1}^{0*} (\td_R \!+\! \td_L)] \nonumber \\
   &- \!\sqrt{2}   c'\! \sin \theta [\hat{d}_M^{0*} s_M^0 (\td_R^*\! - \!\td_L^*) - \hat{d}_{M\!\pm \!1}^{0} s_{M \!\pm \!1}^{0*} (\td_R \!- \!\td_L)] \nonumber \\
   &+ \!2 i \! \cos \theta           [\hat{d}_M^{0*} s_M^0 \td_z^* \!+ \!\hat{d}_{M\!\pm \!1}^{0} s_{M \!\pm \!1}^{0*} \td_z] \nonumber \\
   &+ \!i c'[(\hat{d}_M^{\pm *} s_M^\pm \!+\! \hat{d}_M^{\mp *} s_M^\mp)(\td_R^*\! + \!\td_L^*)\! +\! \hat{d}_{M\!\pm \!2}^\mp s_{M \!\pm \!2}^{\mp *} (\td_R\! +\! \td_L)] \nonumber \\
   &+ \!s'\![(\hat{d}_M^{\pm *} s_M^\pm \!+\! \hat{d}_M^{\mp *} s_M^\mp)(\td_R^* \!-\! \td_L^*)\! - \!\hat{d}_{M\!\pm \!2}^\mp s_{M\! \pm \!2}^{\mp *} (\td_R \!- \!\td_L)] \nonumber \\
   &\mp \!ic' \!\cos \theta [(\hat{d}_M^{\pm *} s_M^\pm \!-\! \hat{d}_M^{\mp *} s_M^\mp)(\td_R^* \!-\! \td_L^*)\! -\! \hat{d}_{M\!\pm \!2}^\mp s_{M \!\pm \!2}^{\mp *} (\td_R\! -\! \td_L)]
        \nonumber \\
   &\mp \!s' \!\cos \theta [(\hat{d}_M^{\pm *} s_M^\pm \!-\! \hat{d}_M^{\mp *} s_M^\mp)(\td_R^*\! + \!\td_L^*)\! -\! \hat{d}_{M\!\pm \!2}^\mp s_{M\! \pm \!2}^{\mp *} (\td_R\! +\! \td_L)]
       \nonumber \\
   &\pm \!\sqrt{2} \!\sin \theta [(\hat{d}_M^{\pm *} s_M^\pm \!-\! \hat{d}_M^{\mp *} s_M^\mp) \td_z^* \!+\! \hat{d}_{M\pm 2}^\mp s_{M\! \pm \!2}^{\mp *} \td_z ] \} \label{eq_invtime_sig}\\
\!\!\!&D_t \Delta_M^0  = P_M^0 \phi(v) - \Gamma_M^0 \Delta_M^0 -\frac{1}{4\hbar}  \Im \{  \nonumber \\
& \!2\sqrt{2}i \hat{d}_M^{0*} s_M^0 [s' \!\sin \theta (\td_R^* \!+ \!\td_L^*) \!+\!i c'\! \sin \theta (\td_R^* \!-\! \td_L^*) \!+ \!\sqrt{2}\td_z^* \cos \theta]
       \nonumber \\
  &\!\!\!+\!\!ic' \![\!(\hat{d}_M^{+*} \!s_M^+ \!\!+\! \hat{d}_M^{-*} \!s_M^-\!)\!(\td_R^* \!\!+\! \td_L^*)\! 
\!+\!\! (\hat{d}_{M\!-\!1}^+ \!s_{M\!-\!1}^{+*} \!\!+\! \hat{d}_{M\!+\!1}^- \!s_{M\!+\!1}^{-*}\!)\!(\td_R\!\! +\! \td_L \!)\!]
       \nonumber \\
  &\!\!\!+\!\!s' \![\!(\hat{d}_M^{+*} \!s_M^+ \!\!+\! \hat{d}_M^{-*} \!s_M^-\!)\!(\td_R^*\!\! - \!\td_L^*)\!
\! -\!\! (\hat{d}_{M\!-\!1}^+ \!s_{M\!-\!1}^{+*}\!\! +\! \hat{d}_{M\!+\!1}^- \!s_{M\!+\!1}^{-*}\!)\!(\td_R \!\!- \!\td_L \!)\!]
       \nonumber \\
  &\!\!\!-\!\!s' \!\!\cos \!\theta [\!(\!\hat{d}_M^{+*} \!s_M^+ \!\!\!-\! \hat{d}_M^{-*} \!s_M^-\!)\!(\!\td_{\!R}^* \!\!+\! \td_{\!L}^*\!) \!
\! -\!\! (\!\hat{d}_{M\!-\!1}^+ \!s_{M\!-\!1}^{+*}\!\! \!-\! \hat{d}_{M\!+\!1}^- \!s_{M\!+\!1}^{-*}\!)\!(\!\td_R \!\!+\! \td_L\!)\!]
       \nonumber \\
  &\!\!\!-\!\!ic' \!\!\cos \!\theta [\!(\!\hat{d}_M^{+*} \!s_M^+ \!\!\!-\! \hat{d}_M^{-*} \!s_M^-\!)\!(\!\td_{\!R}^* \!\!-\! \td_{\!L}^*\!) \!
\!+\!\! (\!\hat{d}_{M\!-\!1}^+ \!s_{M\!-\!1}^{+*}\!\!\!-\! \hat{d}_{M\!+\!1}^- \!s_{M\!+\!1}^{-*}\!)\!(\!\td_{\!R}\! \!-\! \td_{\!L}\!)\!]
       \nonumber \\
  &\!\!\!+\!\!\sqrt{2}\!\sin \!\theta [\!(\hat{d}_M^{+*} \!s_M^+ \!\!-\! \hat{d}_M^{-*} \!s_M^- \!)\td_z^* \!
\!-\!\! (\!\hat{d}_{M\!-\!1}^+ \!s_{M\!-\!1}^{+*} \!\!-\! \hat{d}_{M\!+\!1}^- \!s_{M\!+\!1}^{-*}\!)\td_z^*\!] \}
\end{align}
\end{subequations}

\subsection{Radiative Transfer Equations}
\label{ss:timert}

In converting the general multi-level radiative transfer equation, eq.(\ref{eq_rt3})
to forms specific to the propagation of a Zeeman group, two problems need to
br overcome: the membership of the sums, and the representation of the dipole
elements in the radiation-based coordinates.

The outer sum is over upper energy levels. 
We may immediately discard all the levels in $J'$ in our Zeeman group, since any
downward electric dipole-allowed transition must begin in $J$ (see Fig.~\ref{fig_trantypes}). The
outer sum therefore encompasses the sub-levels of $J$: it is over
$M$ from $-J$ to $+J$. The inner sum is over
the lower levels of the electric dipole transitions, and is therefore immediately
limited to the sub-levels of $J'$. However, again with reference to Fig.~\ref{fig_trantypes}, we
can see that for a given $M$ in $J$ there are at most three possible lower levels,
each corresponding to a transition of different helical type because of the
selection rule $\Delta M = 0,\pm1$. Assuming that all of these transitions exist
for a given $M$, we write the inner sum explicitly, so exchanging the order of
summation and (velocity) integration we obtain from eq.(\ref{eq_rt3}) the Zeeman group
transfer equation,
\begin{equation}
d_t \td_q
=
\frac{\omega_0}{2\epsilon_0} \int_{-\infty}^\infty dv \sum_{M=-J}^{J} 
 (\hat{d}_{M,q}^{+*}  s_M^+ +  \hat{d}_{M,q}^{0*}  s_M^0 + \hat{d}_{M,q}^{-*}  s_M^- ) .
\label{eq_rt4}
\end{equation}
where $q=R,L,z$, and functional dependencies have been suppressed for brevity.

To address the second problem, we need a representation of the dipole
vectors of the three transition types in components based on the global axis
system. We already have this: eq.(\ref{eq_dipgold}), but we need to take the
complex conjugate to match all the dipole components in eq.(\ref{eq_rt4}).
Noting that the spherical vectors used in eq.(\ref{eq_rldef}) have the
property $\hv{e}_R = \hv{e}_L^*$ and $\hv{e}_L = \hv{e}_R^*$, the correct substitutions can
be identified for all three values of $q$, and generating three versions of
eq.(\ref{eq_rt4}):
\begin{subequations} \label{eq_timert}
\begin{align}
d_t \td_R &= \frac{\omega_0}{4\epsilon_0}\int_{-\infty}^\infty dv \sum_{M=-J}^{J}
 \{
\hat{d}_M^{+*} s_M^+ (1-\cos \theta) (c'-is') \nonumber \\
 &+ \sqrt{2} \hat{d}_M^{0*} s_M^0 \sin \theta (s'+ic')
 + \hat{d}_M^{-*} s_M^- (1+\cos \theta) (c'-is') \} \label{eq_timert_R} \\
d_t \td_L &= \frac{\omega_0}{4\epsilon_0}\int_{-\infty}^\infty dv \sum_{M=-J}^{J}
\{
\hat{d}_M^{+*} s_M^+ (1+\cos \theta) (c'+is') \nonumber \\
 &+ \sqrt{2} \hat{d}_M^{0*} s_M^0 \sin \theta (s'-ic')
 + \hat{d}_M^{-*} s_M^- (1-\cos \theta) (c'+is') \} \label{eq_timert_L} \\
d_t \td_z &= \frac{\omega_0}{4\epsilon_0}\int_{-\infty}^\infty dv \sum_{M=-J}^{J}
\{
\hat{d}_M^{+*} s_M^+ (-\sqrt{2}i \sin \theta ) \nonumber \\
 &+ 2 \hat{d}_M^{0*} s_M^0 \cos \theta 
 + \hat{d}_M^{-*} s_M^- (\sqrt{2}i \sin \theta ) \} \label{eq_timert_z} 
\end{align}
\end{subequations}

The groups of equations, eq.(\ref{eq_offtime}), eq.(\ref{eq_invtime}) and eq.(\ref{eq_timert}) now form a complete set
of governing equations for the solution of the OAM maser problem with full polarization in the
time domain.
A reduction of the maser governing equations to the more
standard Zeeman system with polarized radiation, but no OAM, may be
effected by setting $c'=1$, $s'=0$ and $\td_z = 0$ in all these
equations.

\section{Frequency-Domain Equations}
\label{s:fdom}

The easiest of the governing equations to transform to the frequency domain
are the radiative transfer equations, because they are linear in the electric
field amplitudes and DM elements. We use eq.(\ref{eq_ffwd}) to write the
time-domain quantities as transforms of the frequency-domain versions. After
differentiation of the transform expression on the left-hand side, only a
$z$-derivative remains, so that a set of PDEs in the time domain has been reduced
to ODEs in frequency \citep{1978PhRvA..17..701M,2009MNRAS.399.1495D}. Formal
inverse transformation via eq.(\ref{eq_finv}) then results in,
\begin{subequations} \label{eq_freqrt}
\begin{align}
\! &\frac{d\td_{R,n}}{dz} = \frac{\omega_0}{4c\epsilon_0}\int_{-\infty}^\infty dv \sum_{M=-J}^{J}
 \{
\hat{d}_M^{+*} s_{M,n}^+ (1-\cos \theta) (c'-is') \nonumber \\
 &+ \sqrt{2} \hat{d}_M^{0*} s_{M,n}^0 \sin \theta (s'+ic')
 + \hat{d}_M^{-*} s_{M,n}^- (1+\cos \theta) (c'-is') \} \label{eq_timert_R} \\
& \frac{d\td_{L,n}}{dz} = \frac{\omega_0}{4\epsilon_0}\int_{-\infty}^\infty dv \sum_{M=-J}^{J}
\{
\hat{d}_M^{+*} s_{M,n}^+ (1+\cos \theta) (c'+is') \nonumber \\
 &+ \sqrt{2} \hat{d}_M^{0*} s_{M,n}^0 \sin \theta (s'-ic')
 + \hat{d}_M^{-*} s_{M,n}^- (1-\cos \theta) (c'+is') \} \label{eq_timert_L} \\
& \frac{d\td_{z,n}}{dz} = \frac{\omega_0}{4\epsilon_0}\int_{-\infty}^\infty dv \sum_{M=-J}^{J}
\{
\hat{d}_M^{+*} s_{M,n}^+ (-\sqrt{2}i \sin \theta ) \nonumber \\
 &+ 2 \hat{d}_M^{0*} s_{M,n}^0 \cos \theta 
 + \hat{d}_M^{-*} s_{M,n}^- (\sqrt{2}i \sin \theta ) \} \label{eq_timert_z} 
\end{align}
\end{subequations}
where we have introduced the following shorthand notations for the Fourier
components:
\begin{subequations} \label{eq_fshort}
\begin{align}
\td_{q,n} &= \td_q(r,\phi,z,\varpi_n) \label{eq_fshort_E} \\
s_{M,n} &= s_M(r,\phi,z,\varpi_n,v) \label{eq_fshort_s} \\
\Delta_{M,n} &= \Delta_M(r,\phi,z,\varpi_n,v) \label{eq_fshort_inv} .
\end{align}
\end{subequations}

Transformation of eq.(\ref{eq_offtime}) is somewhat more difficult because of
the appearance of products of inversions and electric field components on the
right-hand sides. These products in the time domain will appear as convolutions
in the frequency domain \citep{1978PhRvA..17..701M,2009MNRAS.396.2319D}. As an
example, we write,
\begin{equation}
\td_R \Delta_M^+ = {\cal F}^{-1} [\td_{R,n}] {\cal F}^{-1}[\Delta_{M,n}^+] = {\cal F}^{-1} [\td_{R} \otimes \Delta_{M}^+]_{n-m} ,
\label{eq_convol}
\end{equation}
where the $\otimes$ symbol denotes the convolution operation. For a continuous
frequency variable, the convolution would be an integral, but for the present
system of finite-width Fourier components, we replace it with the sum,
\begin{equation}
[\td_{R} \otimes \Delta_{M}^+]_{n-m} = (2\pi)^{-1} \sum_{m=-\infty}^\infty \td_{R,m} \Delta_{M,n-m}^+ ,
\label{eq_convolsum}
\end{equation}
and similar expressions for other polarizations and transition types. With the
help of eq.(\ref{eq_convol}) and eq.(\ref{eq_convolsum}), and dropping terms
of order $v/c$ in size, the transformed versions of the evolution equations for
off-diagonal elements of the DM become the algebraic equations:
\begin{subequations} \label{eq_freqoff}
\begin{align}
s_{M,n}^\pm &= \frac{\hat{d}_M^\pm \tilde{L}_M^\pm}{2 \hbar }\sum_{m=-\infty}^\infty \Delta_{M,n-m}^\pm \{(1\mp \cos \theta)(c'+is') \td_{R,m} \nonumber \\
          &+ (1 \pm \cos \theta ) (c'-is') \td_{L,m} \pm \sqrt{2} i \sin \theta \td_{z,m} 
\} \label{eq_freqoff_sigma} \\
s_{M,n}^0 &= \frac{\hat{d}_M^0 \tilde{L}_M^0}{\sqrt{2} \hbar }\sum_{m=-\infty}^\infty \Delta_{M,n-m}^0 \{ \sin \theta (s'-ic') \td_{R,m} \nonumber \\
          &+ \sin \theta (s'+ic') \td_{L,m} + \sqrt{2} \cos \theta \td_{z,m}
\} \label{eq_freqoff_pi}
\end{align}                                               
\end{subequations}
where $\tilde{L}_M^\pm, \tilde{L}_M^0$ are the complex Lorentzian functions, defined by
\begin{equation}
\tilde{L}_M^{\pm ,0} = \frac{1}
                     {2\pi [\gamma^{\pm ,0} -i (\varpi_n - \Delta \omega_M^{\pm ,0} - v \omega_0 /c)]} ,
\label{eq_complor}
\end{equation}
where the optional symbol $\pm ,0$ as a superscript encompasses all three transition types.
The normalised real Lorentzian is, with the constant of $2\pi$ as used in eq.(\ref{eq_complor}),
\begin{equation}
L_M^{\pm ,0} = \tilde{L}_M^{\pm , 0} + \tilde{L}_M^{(\pm, 0)*} = \frac{\gamma_M^{\pm ,0} /\pi}
                             {(\gamma_M^{\pm, 0} )^2 + (\varpi_n - \Delta \omega_M^{\pm ,0} - v \omega_0 /c)^2} .
\label{eq_lor}
\end{equation}

It is also necessary to transform the inversion equation, eq.(\ref{eq_invtime}), and this
also contains time-domain products that will transform to frequency-domain convolutions.
In this case, the products are all of electric field complex amplitudes with off-diagonal
DM elements. Operations on the left-hand side proceed in a similar manner to those
for the off-diagonal elements of the DM: terms of order $v/c$ in the Doppler velocity
are ignored, and the transformd equations are algebraic. On the right-hand side, there are
two complicating issues: the first is the pumping term that is not a function of time. We
write this as $P_M^+ \phi (v) {\cal F}^{-1} [\delta_n ]$, where we have taken the $\sigma^+$
version of eq.(\ref{eq_invtime_sig}) as an example, and $\delta_n$ is a $\delta$-function that
has the value $1$ for $n=0$ and zero for any other Fourier component. The second complication
is that we must now transform complex conjugate quantities. We assume here that the
discrete-width transforms used here follow the usual rules, that is for example,
\begin{equation}
\td_R^*(t) = ({\cal F}^{-1}[\td_{R,n}])^* = {\cal F}^{-1}[\td_{R,-n}^*] .
\label{eq_ftconj}
\end{equation}
The products that will transform to convolutions are either of a conjugate complex
amplitude and an ordinary off-diagonal DM element or vice-versa, resulting in the
discrete representations,
\begin{subequations} \label{eq_convulconj}
\begin{align}
\! & {\cal F}^{-1}[\td_{R,n}] {\cal F}^{-1}[s_{M+1,-n}^{0*}] = {\cal F}^{-1}[\td_R \otimes s_{M+1}^{0*}]_{n-m} \nonumber \\
&= (2\pi)^{-1} \sum_{m=-\infty}^\infty \td_{R,m} s_{M+1,m-n}^{0*} \label{eq_convulconj_a} \\
\! & {\cal F}^{-1}[\td_{R,-n}^*] {\cal F}^{-1}[s_{M+1,n}^0] = {\cal F}^{-1}[\td_R^* \otimes s_{M+1}^{0}]_{n+m} \nonumber \\
& = (2\pi)^{-1} \sum_{m=-\infty}^\infty \td_{R,m}^* s_{M+1,m+n}^0 , \label{eq_convulconj_b}
\end{align}
\end{subequations}
and similarly for other transition types and polarizations. The resulting expressions for
the inversions are
\begin{subequations} \label{eq_freqinv}
\begin{align}
\! & \Delta_{M,n}^\pm = P_M^\pm \phi (v) \delta_n /\Gamma_M^\pm - \tilde{\cal L}_M^\pm /(2 \hbar) \sum_{m=-\infty}^\infty \{ \nonumber \\
& \frac{\sin \theta}{\sqrt{2}} [ (is'-c') (\hat{d}_M^{0*} s_{M,m+n}^0 \td_{R,m}^* + \hat{d}_{M\pm1}^0 s_{M\pm1,m-n}^{0*} \td_{L,m}) \nonumber \\
\;\; &+ (is'+c') (\hat{d}_M^{0*} s_{M,m+n}^0 \td_{L,m}^* + \hat{d}_{M\pm1}^0 s_{M\pm1,m-n}^{0*} \td_{R,m}) \nonumber \\
 &\pm 2\hat{d}_M^{\pm *} s_{M,m\!+\!n}^\pm \td_{z,m}^* \!-\! \hat{d}_M^{\mp *} s_{M,m\!+\!n}^\mp \td_{z,m}^* \!+\! \hat{d}_{M\!\pm \!2}^\mp s_{M\!\pm \!2,m\!-\!n}^{\mp *} \td_{z,m} ] \nonumber \\
&+ \frac{(1\pm \cos \theta)}{2} [ (s'+ic') (\hat{d}_M^{\mp *} s_{M,m+n}^\mp \td_{R,m}^* ) \nonumber \\
&-(s'-ic')( 2 \hat{d}_M^{\pm *} s_{M,m+n}^\pm \td_{L,m}^* + \hat{d}_{M\pm2}^\mp s_{M\pm2,m-n}^{\mp *} \td_{R,m} ) ] \nonumber \\
&+ \frac{(1\mp \cos \theta)}{2} [-(s'-ic') ( \hat{d}_M^{\mp *} s_{M,m+n}^\mp \td_{L,m}^* ) \nonumber \\
&+ (s'+ic') ( 2 \hat{d}_M^{\pm *} s_{M,m+n}^\pm \td_{R,m}^* + \hat{d}_{M\pm2}^\mp s_{M\pm2,m-n}^{\mp *} \td_{L,m} ) ] \nonumber \\
&+i\cos \theta [ \hat{d}_M^{0*} s_{M,m+n}^0 \td_{z,m}^* + \hat{d}_{M\pm1}^0 s_{M\pm1,m-n}^{0*} \td_{z,m}] \}
\label{eq_freqinv_sig} \\
\! & \Delta_{M,n}^0 = P_M^0 \phi (v) \delta_n /\Gamma_M^0 - \tilde{\cal L}_M^0 /(2 \hbar) \sum_{m=-\infty}^\infty \{ \nonumber \\
& \frac{\sin \theta}{\sqrt{2}} [ 2 \hat{d}_M^{0*} s_{M,m+n}^0 ( (is'-c') \td_{R,m}^* + (is'+c') \td_{L,m}^* ) \nonumber \\
&+ (\hat{d}_M^{+*} s_{M,m+n}^+  - \hat{d}_M^{-*} s_{M,m+n}^-  ) \td_{z,m}^* \nonumber \\
&- (\hat{d}_{M-1}^+ s_{M-1,m-n}^{+*}  - \hat{d}_{M+1}^- s_{M+1,m-n}^{-*} )\td_{z,m} ] \nonumber \\
&+\! \frac{(1 \!+\! \cos \theta)}{2} [ (s'\!+\!ic') ( \hat{d}_M^{-*} s_{M,m\!+\!n}^- \td_{R,m}^* \!\!+\! \hat{d}_{M\!-\!1}^+ s_{M\!-\!1,m\!-\!n}^{+*} \td_{L,m} ) \nonumber \\
&- (s'-ic') ( \hat{d}_M^{+*} s_{M,m+n}^+ \td_{L,m}^* + \hat{d}_{M+1}^- s_{M+1,m-n}^{-*} \td_{R,m}) ] \nonumber \\
&+\! \frac{(1 \!- \! \cos \theta)}{2} [ (s'\!+\!ic') ( \hat{d}_M^{+*} s_{M,m\!+\!n}^+ \td_{R,m}^* \!\!+\! \hat{d}_{M\!+\!1}^- s_{M\!+\!1,m\!-\!n}^{-*} \td_{L,m} ) \nonumber \\ 
&- (s'-ic') ( \hat{d}_M^{-*} s_{M,m+n}^- \td_{L,m}^* + \hat{d}_{M-1}^+ s_{M-1,m-n}^{+*} \td_{R,m}) ] \nonumber \\
&+ 2i\cos \theta \hat{d}_M^{0*} s_{M,m+n}^0 \td_{z,m}^* , \label{eq_freqinv_pi} 
\end{align}
\end{subequations}
where optional signs have been used in eq.(\ref{eq_freqinv_sig}) to allow this equation to be used
for both $\sigma^+$ (upper sign) and $\sigma^-$ (lower sign). The complex Lorentzian function
$\tilde{\cal L}_M^{\pm , 0}$ is defined as
\begin{equation}
\tilde{\cal L}_M^{\pm , 0} = (2\pi)^{-1} [\Gamma_M^{\pm , 0} - i \varpi_n ]^{-1} .
\label{eq_Glor}
\end{equation}

The set of equations comprising the subsets eq.(\ref{eq_freqrt}), eq.(\ref{eq_freqoff})
and eq.(\ref{eq_freqinv}) now form a closed set for the solution of the full maser
amplification problem with full polarization plus OAM in the frequency domain.
All processes of saturation, population pulsation and mode coupling, for example,
are included. The only auxilliary equations required are the definitions of the
molecular velocity distribution function, eq.(\ref{eq_gaussian}), and the definitions of the
Zeeman frequency shifts, eq.(\ref{eq_zshift}). Suitable boundary conditions are to inject
radiation for all Fourier components at $z=0$ with no polarization and no
OAM. This amounts to setting all $z$-component complex amplitudes to zero,
and setting the left- and right-handed complex amplitudes to values appropriate
for background noise with Gaussian statistics. A suitable prescription is
to set each phase to an independent value drawn from a uniform distribution, and
each real amplitude to an independent value drawn from a normal distribution.

Such complexity is not necessary to answer the question of whether a maser can
amplify OAM. By analogy with polarization, we only need to know whether it can 
grow: ignore all the more subtle effects, including even saturation, and
study growth of intensity-like variables.

\section{Classical Reduction}
\label{classic}

To make a classical reduction of our system of frequency-domain governing equations,
we make the following approximations: (i) different Fourier components of the radiation
field remain uncorrelated for any degree of saturation; (ii) population is restricted to
a single central Fourier component, numbered zero; (iii) radiation statistics remain
Gaussian at all signal strengths; (iv) Lorentzian functions act effectively
as $\delta$-functions. Point (iii) allows all correlation expressions of
electric field amplitudes to be reduced to correlations of orders 1 and 2 only.

Immediate effects of the above reductions include the collapse of the sums over
$m$ in eq.(\ref{eq_freqoff}) to a single element with
$m=n$. An important consequence is that only the
central Fourier component of the population inversion is required for each
transition type, that is $\Delta_{M,0}^{\pm , 0}$, and similarly for the other subscripts
from $M-2$ to $M+2$.
This restriction on the inversion also leads to a collapse of the sum over $m$ in
eq.(\ref{eq_freqinv}), but we do not need to consider this further because we are
considering only unsaturated amplification. The radiative transfer equations are
not immediately affected by the classical approximations.

\subsection{Stokes and OAM Parameters}
\label{ss:stokes}

A vector of four Stokes parameters gives a complete description of polarized
radiation, but is inadequate to address the additional complication of radiation with OAM.
Since we are measuring the amount of OAM by the presence of a $z$-component of the electric
field, it makes sense to define new parameters that measure both the overall intensity
present in this $z$-component and the intensity of its interaction with the left- and
right-handed polarizations. To this end, we define the OAM parameters,
\begin{subequations} \label{eq_oparms}
\begin{align}
J_n &= \langle {\cal J}_n \rangle = \langle \td_{z,n} \td_{z,n}^* \rangle \label{eq_oparams_J} \\
G_n &=  \langle {\cal G}_n \rangle = \langle \td_{z,n} \td_{R,n}^* + \td_{R,n} \td_{z,n}^* \rangle \label{eq_oparms_G} \\
H_n &= \langle {\cal H}_n \rangle = \langle \td_{z,n} \td_{L,n}^* + \td_{L,n} \td_{z,n}^* \rangle \label{eq_oparms_H} \\
W_n &=  \langle {\cal W}_n \rangle = i \langle \td_{z,n} \td_{R,n}^* - \td_{R,n} \td_{z,n}^* \rangle \label{eq_oparms_W} \\
X_n &= \langle {\cal X}_n \rangle = i \langle \td_{z,n} \td_{L,n}^* - \td_{L,n} \td_{z,n}^* \rangle \label{eq_oparms_X}
\end{align}
\end{subequations}
noting that, like the Stokes parameters, all these new parameters are real.
The angle brackets denote an average over a good statistical number of
realizations of the electric field. Each realization, in the present context,
corresponds to an individual spectral sample of duration $T$ \citep{1978PhRvA..17..701M,2009MNRAS.396.2319D}.
The first of these new parameters is a measure of the total intensity of
OAM present, as measured from the intensity of the $z$-component of the
electric field. The other four parameters are a measure of the interaction of
the OAM with the conventional electric field components in the $xy$ (or $r\phi$)
plane. By summing the squares of eq.(\ref{eq_oparms_G})-eq.(\ref{eq_oparms_X}), it
is straightforward to show that these parameters satisfy the relation,
\begin{equation}
G_n^2 + H_n^2 + W_n^2 + X_n^2 = 2 I_n J_n .
\label{eq_sqoparms}
\end{equation}

We also maintain the definition of the Stokes parameters that is compatible
with the IEEE convention on left- and right-handed polarization, the IAU axis
system, and the IAU convention that defines positive Stokes-$V$ as an excess
of right-handed over left-handed polarization. Such a set is,
\begin{subequations} \label{eq_sparms}
\begin{align}
I_n &= \langle {\cal I}_n \rangle = \langle \td_{R,n} \td_{R,n}^* + \td_{L,n} \td_{L,n}^* \rangle \label{eq_sparms_I} \\
Q_n &= \langle {\cal Q}_n \rangle = \langle \td_{R,n} \td_{L,n}^* + \td_{L,n} \td_{R,n}^* \rangle \label{eq_sparms_Q} \\
U_n &= \langle {\cal U}_n \rangle = i \langle \td_{R,n} \td_{L,n}^* - \td_{L,n} \td_{R,n}^* \rangle \label{eq_sparms_U} \\
V_n &= \langle {\cal V}_n \rangle = \langle \td_{R,n} \td_{R,n}^* - \td_{L,n} \td_{L,n}^* \rangle \label{eq_sparms_V} .
\end{align}
\end{subequations}
We note that these Stokes parameters, and the OAM parameters in eq.(\ref{eq_oparms}), should formally
be multiplied by a constant that gives them units of specific intensity. However, as we will construct
equations below, in Section~\ref{ss:intense}, that are linear in the Stokes and OAM parameters, we
omit this constant as it will cancel from both sides of each equation.

\subsection{Intensity Equations}
\label{ss:intense}

We construct equations for the transport of the OAM and Stokes parameters by differentiating the definitions
in eq.(\ref{eq_oparms}) and eq.(\ref{eq_sparms}) with respect to $z$. For example, the unaveraged form of the parameter ${\cal G}_n$
from eq.(\ref{eq_oparms_G}) gives us,
\begin{equation}
\frac{d{\cal G}_n}{dz} = \td_{z,n} \frac{d\td_{R,n}^*}{dz} + \td_{R,n}^* \frac{d\td_{z,n} }{dz}
                       + \td_{R,n} \frac{\td_{z,n}^*}{dz} +  \td_{z,n}^* \frac{\td_{R,n} }{dz} .
\label{eq_gderiv}
\end{equation}
The right-hand sides may be constructed from eq.(\ref{eq_timert_R}), eq.(\ref{eq_timert_z}) and
their complex conjugates. The equations contain various forms of off-diagonal DM element, but
these can in turn be eliminated via eq.(\ref{eq_freqoff}), noting that, in our classical
approximation, $m=n$ is the only term in the sum, and all inversions revert to the central
Fourier component. A final realization average of eq.(\ref{eq_gderiv}) yields the 
power-spectrum form, $G_n$. The same method may also be employed for all the other Stokes
and OAM parameters. The results are summarized in the following five transfer equations for
the OAM parameters, as defined in eq.(\ref{eq_oparms})
\begin{subequations} \label{eq_parmgold}
\begin{align}
\! & \frac{dJ_n}{dz} = \frac{1}{8\epsilon_0 \hbar} \sum_{M=-J}^J \left\{
 2 J_n (2D_{M,n}^0 - \Pi_{M,n} \sin^2 \theta ) \right. \nonumber \\
 &\left. + \frac{\sin \theta}{\sqrt{2}} (R_{M,n} + \Pi_{M,n} \cos \theta) (s' G_n + c' W_n ) \right. \nonumber \\
 &\left. + \frac{\sin \theta}{\sqrt{2}} (\Pi_{M,n} \cos \theta - R_{M,n}) (s' H_n - c' X_n ) 
                                                         \right\} \label{eq_parmgold_J} \\
& \frac{dG_n}{dz} = \frac{1}{8\epsilon_0 \hbar} \sum_{M=-J}^J \left\{
G_n (T_{M,n} -\frac{1}{2} \Pi_{M,n} \sin^2 \theta - R_{M,n} \cos \theta ) \right. \nonumber \\
&\left. + \frac{s'\sin \theta}{\sqrt{2}}(R_{M,n} + \Pi_{M,n} \cos \theta)(I_n + V_n + 2J_n) \right. \nonumber \\
&\left. - \frac{\sin \theta}{\sqrt{2}}(R_{M,n} - \Pi_{M,n} \cos \theta)(s'Q_n - c'U_n) \right. \nonumber \\
&\left. - \frac{\Pi_{M,n}\sin^2 \theta}{2} (H_n \cos 2\phi' + X_n \sin 2\phi' ) \right\} \\
& \frac{dH_n}{dz} = \frac{1}{8\epsilon_0 \hbar} \sum_{M=-J}^J \left\{
H_n (T_{M,n} -\frac{1}{2} \Pi_{M,n} \sin^2 \theta + R_{M,n} \cos \theta ) \right. \nonumber \\
&\left. + \frac{s'\sin \theta}{\sqrt{2}}(\Pi_{M,n} \cos \theta - R_{M,n})(I_n - V_n + 2J_n) \right. \nonumber \\
&\left. + \frac{\sin \theta}{\sqrt{2}}(R_{M,n} + \Pi_{M,n} \cos \theta)(s'Q_n - c'U_n) \right. \nonumber \\
&\left. - \frac{\Pi_{M,n}\sin^2 \theta}{2} (G_n \cos 2\phi' - W_n \sin 2\phi' ) \right\} \\
&\frac{dW_n}{dz} = \frac{1}{8\epsilon_0 \hbar} \sum_{M=-J}^J \left\{
W_n (T_{M,n} -\frac{1}{2} \Pi_{M,n} \sin^2 \theta - R_{M,n} \cos \theta ) \right. \nonumber \\
&\left. + \frac{c'\sin \theta}{\sqrt{2}}(R_{M,n} + \Pi_{M,n} \cos \theta )(I_n + V_n + 2J_n) \right. \nonumber \\
&\left. + \frac{\sin \theta}{\sqrt{2}}(R_{M,n} - \Pi_{M,n} \cos \theta)(c'Q_n + s'U_n) \right. \nonumber \\
&\left. - \frac{\Pi_{M,n}\sin^2 \theta}{2} (X_n \cos 2\phi' - H_n \sin 2\phi' ) \right\} \\
&\frac{dX_n}{dz} = \frac{1}{8\epsilon_0 \hbar} \sum_{M=-J}^J \left\{
X_n (T_{M,n} -\frac{1}{2} \Pi_{M,n} \sin^2 \theta + R_{M,n} \cos \theta ) \right. \nonumber \\
&\left. + \frac{c'\sin \theta}{\sqrt{2}}(R_{M,n} - \Pi_{M,n} \cos \theta )(I_n - V_n + 2J_n) \right. \nonumber \\
&\left. + \frac{\sin \theta}{\sqrt{2}}(R_{M,n} + \Pi_{M,n} \cos \theta)(c'Q_n + s'U_n) \right. \nonumber \\
&\left. - \frac{\Pi_{M,n}\sin^2 \theta}{2} (W_n \cos 2\phi' + G_n \sin 2\phi' ) \right\},
\end{align}
\end{subequations}
and in the four following transfer equations for the Stokes parameters, as defined in eq.(\ref{eq_sparms}),
noting that these and the five parts of eq.(\ref{eq_oparms}) together form a coupled set of nine ODEs.
\begin{subequations} \label{eq_stokesgold}
\begin{align}
\! &\frac{dI_n}{dz} = \frac{1}{8\epsilon_0 \hbar} \sum_{M=-J}^J \left\{
2I_n (D_{M,n}^+ + D_{M,n}^-) -2 R_{M,n} V_n \cos \theta \right. \nonumber \\
&\left. + \Pi_{M,n} (I_n -Q_n \cos 2\phi' - U_n \sin 2\phi') \sin^2 \theta \right. \nonumber \\
&\left. + \frac{\sin \theta}{\sqrt{2}}(R_{M,n} + \Pi_{M,n} \cos \theta )(s'G_n + c'W_n) \right. \nonumber \\
&\left. - \frac{\sin \theta}{\sqrt{2}}(R_{M,n} - \Pi_{M,n} \cos \theta )(s'H_n - c'X_n) \right\} \\
&\frac{dV_n}{dz} = \frac{1}{8\epsilon_0 \hbar} \sum_{M=-J}^J \left\{
[2(D_{M,n}^+ + D_{M,n}^-) + \Pi_{M,n}\sin^2 \theta ] V_n \right. \nonumber \\
&\left. - 2 R_{M,n} I_n \cos \theta + \frac{\sin \theta}{\sqrt{2}}(R_{M,n} - \Pi_{M,n} \cos \theta )(s'H_n - c'X_n) \right. \nonumber \\
&\left. + \frac{\sin \theta}{\sqrt{2}}(R_{M,n} + \Pi_{M,n} \cos \theta )(s'G_n + c'W_n) \right\} \\
&\frac{dQ_n}{dz} = \frac{1}{8\epsilon_0 \hbar} \sum_{M=-J}^J \left\{
[ 2(D_{M,n}^+ + D_{M,n}^-) + \Pi_{M,n} \sin^2 \theta] Q_n  \right. \nonumber \\
&\left. + [c' (W_n + X_n) - s' (G_n - H_n) ]\frac{\sin \theta}{\sqrt{2}} (R_{M,n}-\Pi_{M,n}\cos \theta) \right. \nonumber \\
&\left. - \Pi_{M,n} I_n \sin^2 \theta \cos 2\phi' \right\} \\
&\frac{dU_n}{dz} = \frac{1}{8\epsilon_0 \hbar} \sum_{M=-J}^J \left\{
[ 2(D_{M,n}^+ + D_{M,n}^-) + \Pi_{M,n} \sin^2 \theta] U_n  \right. \nonumber \\
&\left. + [c' (G_n - H_n) + s' (W_n + X_n) ]\frac{\sin \theta}{\sqrt{2}} (R_{M,n}-\Pi_{M,n}\cos \theta) \right. \nonumber \\
&\left. - \Pi_{M,n} I_n \sin^2 \theta \sin 2\phi' \right\} .
\end{align}
\end{subequations}

In deriving the set of equations, eq.(\ref{eq_parmgold}), we have defined the following
variables related to the inversions:
\begin{subequations} \label{eq_dshorty}
\begin{align}
\Pi_{M,n} &= 2D_{M,n}^0 - D_{M,n}^+ - D_{M,n}^- \label{eq_dshorty_pi} \\
R_{M,n} &= D_{M,n}^+ - D_{M,n}^- \label{eq_dshorty_R} \\
T_{M,n} &=  2D_{M,n}^0 + D_{M,n}^+ + D_{M,n}^- \label{eq_dshorty_T}
\end{align}
\end{subequations}
The first of these variables, $\Pi_{M,n}$, represents any dominance of the inversion
in the $\pi$-transition over the summed inversions in both $\sigma$-transitions at
a frequency local to Fourier component $n$.
The second variable, $R_{M,n}$, expresses any imbalance in the inversions of the 
$\sigma$-transitions at the same frequency. This imbalance will be large in a Zeeman group such as
those found in typical OH masers where, in a magnetic field of a few mG, the
$\sigma^+$ and $\sigma-$ spectral lines propagate effectively independently.

The inversion-related functions $D_{M,n}^{\pm , 0}$ are defined in general through the
velocity integrals,
\begin{equation}
D_{M,n}^{\pm , 0} =(\omega_0 /c)|\hat{d}_M^{\pm , 0}|^2 \left\langle \int_{-\infty}^\infty \Delta_{M,0}^{\pm , 0} (v) L_M^{\pm , 0} (v) dv \right\rangle,
\label{eq_bigD}
\end{equation}
where the spatial functional dependence of the inversion has been omitted. The factor of
$\omega_0 /c$ is included because the Lorentzian, as defined in eq.(\ref{eq_lor}), is normalized in
frequency, and this extra factor is required to normalize it in velocity. In the
specific case of negligible saturation, all terms in the radiation complex amplitudes
may be dropped from eq.(\ref{eq_freqinv}), so that, in all three versions, the inversion
in Fourier component zero reduces to the combined form,
\begin{equation}
\Delta_{M,0}^{\pm , 0} (v) = P_M^{\pm ,0} \phi(v) / \Gamma_M^{\pm ,0} ,
\label{eq_unsatinv}
\end{equation} 
and, when this is substituted into eq.(\ref{eq_bigD}), the integration can be carried
out, on the assumption that the Lorentzian tends to a $\delta$-function. The realization
averaging can also be dropped, because eq.(\ref{eq_unsatinv}) contains no electric field
amplitudes or elements of the DM. With these points noted,
\begin{equation}
D_{M,n}^{\pm , 0} =(|\hat{d}_M^{\pm , 0}|^2 P_M^{\pm , 0} /\Gamma_M^{\pm ,0} ) \phi [ (c/\omega_0)(\varpi_n - \Delta \omega_M^{\pm ,0})] .
\label{eq_bigDint}
\end{equation}
Note that the Gaussian function is now in terms of local frequency, and peaks at the Zeeman-shifted response frequency
of the relevant transition type.

\section{Discussion}
\label{discuss}

Several useful points may be raised by considering various limits and
special cases of eq.(\ref{eq_parmgold}) and eq.(\ref{eq_stokesgold}). The first is
that the parameter $J_n$ is not directly coupled to any of the Stokes parameters, as
may be seen from eq.(\ref{eq_parmgold_J}). As this parameter effectively represents
the intensity of the $z$-component of the electric field, such intensity can only
arise indirectly from the Stokes parameters via the other four OAM parameters that
represent interactions of the $z$-component of the electric field with circularly-polarized
amplitudes in the $x-y$ plane. However, eq.(\ref{eq_parmgold_J}) does contain a
term in $J_n$ on the right-hand side that allows it to amplify itself, even if it
needs $G_n,H_n,W_n$ and $X_n$ and the Stokes parameters to grow from an initally
zero background level.

The second main point is that any effect involving OAM, at least if we require it
to grow from an OAM-free background, will be strongest when the magnetic field is
close to perpendicular to the $z$-axis ($\sin \theta = 1$ or $\theta = \pi/2$).
If we take this limit, then $J_n$ remains coupled to $G_n,H_n,W_n$ and $X_n$, whilst
these latter four parameters remain coupled to all four Stokes parameters as well
as to linear combinations of themselves with coefficients formed from certain trigonometric functions
of $\phi'$. The equations describing the evolution of the Stokes parameters retain
a similar coupling to the OAM parameters, though not directly to $J_n$ as discussed
above. By contrast, in the other limit, where the magnetic field lies parallel to
the $z$-axis, most of the above couplings are eliminated: the Stokes parameters couple
only to themselves and combinations of other Stokes parameters. All five OAM parameters
are left with the capacity only to amplify themselves, so no OAM at all can result
from an OAM-free background in the parallel configuration.

A third important point, when considering the case where $\sin \theta = 1$, is
that all the terms that couple Stokes parameters to OAM parameters include
a factor of $R_{M,n}$. From eq.(\ref{eq_dshorty_R}), we can see that this parameter
is essentially the difference between the inversions in the $\sigma^+$ and $\sigma^-$
transitions. Unless we have a pumping process that is asymmetric, favouring one
of these over the other (say $P_M^+ > P_M^-$), then we will obtain a number very
close to zero for typical closed shell maser molecules, such as water, methanol
and SiO, where the Zeeman splitting is likely to be dwarfed by the inhomogeneous
(Doppler) line width. The most likely source of OAM is therefore in OH (and possibly CH) masers,
where the Zeeman splitting may comfortably exceed the Doppler width. In this case,
if we choose a Fourier component, $n$, near the centre of the $\sigma^+$ line, we
will find $R_{M,n} \simeq D_{M,n}^+$, with almost no contamination from the $\sigma^-$
line, even if the pumping is symmetric. The best possible situation for generating
OAM in a maser, from an OAM-free background, is therefore in an open-shell molecule
with a large Zeeman splitting, from a $\sigma$-type transition and a non-uniform magnetic field lying in
the $x-y$ plane. Under these circumstances, the OAM parameters $G_n,H_n,W_n$ and $X_n$
can interact directly with Stokes-$I$ and (independently) with $J_n$. Note that
the terms that couple $G_n,H_n,W_n$ and $X_n$ to $J_n$ also contain a comman factor
of $R_{M,n}$ when $\theta = \pi /2$. Generation of OAM may offer a new
explanation for the observation that linearly polarized Zeeman triplets, expected
for a perpendicular magnetic field, are very rare or absent.

\subsection{Restricted Solutions}
\label{ss:restricted}

Having made the general observations above, we now consider analytic solutions
of eq.(\ref{eq_parmgold}) and eq.(\ref{eq_stokesgold}) under a set of conditions that correspond to optimum generation
of OAM: we take the magnetic field to be perpendicular to the line of sight
($\theta=\pi/2$), and pick a Fourier component, $n=k$, that corresponds to the line
centre of the $\sigma^+$ line in a molecule and transition (for example the
rotational ground state of OH) that has a Zeeman splitting that can easily
exceed the Doppler line width. Under these conditions, we may make the
reductions, $D_{M,k}^- = 0$, $R_{M,k} = T_{M,k}=D_{M,k}$ and $\Pi_{M,k}=-D_{M,k}$, so
that $D_{M,k}^+$ becomes a common factor on the right-hand side of all the governing
equations. We extract this factor, and use it to construct the dimensionless line-of-sight
distance,
\begin{equation}
d\zeta = dz D_{M,k}^+ /(8\epsilon_0 \hbar) .
\label{eq_zetadef}
\end{equation}
We assume that the seed radiation of the maser is unpolarized and OAM-free, with
Stokes-$I$ intensity equal to $I_{BG}$. We divide all the governing equations by
$I_{BG}$, and define dimensionless Stokes and OAM parameters as, for example,
$i_k = I_k /I_{BG}$, $q_k = Q_k /I_{BG}$, and similarly for all the other parameters,
writing lower-case letters for the dimensionless forms.

We consider the magnetic field to be of ideal quadrupole form, with Cartesian
components given by $B_x = \vec{B}_0 \cdot \vec{y}$ and $B_y = \vec{B}_0 \cdot \vec{x}$,
where $\vec{B}_0$ is a constant. We pick, for the sake of example, a point on the
$y$-axis, where $\phi=\pi/2$. At this point, $B_y=0$ and $B_x = B_0 y$, and from
eq.(\ref{eq_defphiprime}), we recover $\phi' = \arccos (0) = \pi /2$. We may
therefore insert into all the governing equations the special values,
$\sin \phi' = s' = 1$, $\cos \phi' = c' = 0$, $\sin 2\phi' = 0$ and
$\cos 2 \phi' = -1$. When this has been done, then by inspection, the nine
governing equations break into two decoupled sets: one set of three contains exclusively
functions of $q_k,w_k,x_k$, whilst the other six equations are entirely free of
these parameters. As all of $q_k,w_k,x_k$ are zero at $z = \zeta = 0$, and are
decoupled from $i_k$, the only parameter with a non-zero background, we discard
the smaller set, and write down the remaining six equations:
\begin{subequations} \label{eq_slicksix}
\begin{align}
dj_k/d\zeta &= 2j_k + (g_k - h_k) /\sqrt{2} \label{eq_slicksix_j} \\
dg_k/d\zeta &= (i_k + v_k + 2 j_k - q_k)/\sqrt{2} + (3 g_k - h_k)/2 \label{eq_slicksix_g} \\
dh_k/d\zeta &= (v_k -i_k - 2 j_k + q_k)/\sqrt{2} + (3 h_k - g_k)/2 \label{eq_slicksix_h} \\
di_k/d\zeta &= i_k - q_k + (g_k - h_k) /\sqrt{2} \label{eq_slicksix_i} \\
dq_k/d\zeta &= q_k - i_k - (g_k - h_k) /\sqrt{2} \label{eq_slicksix_q} \\
dv_k/d\zeta &= v_k + (g_k + h_k) /\sqrt{2} . \label{eq_slicksix_v} 
\end{align}
\end{subequations}

\begin{figure}
\includegraphics[width=130mm]{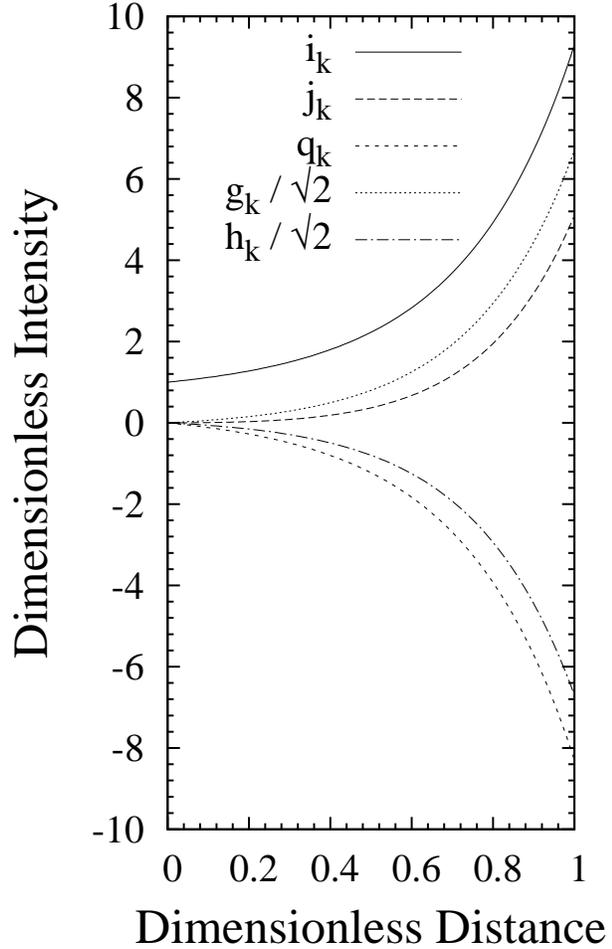}
\caption{The non-zero dimensionless Stokes and OAM parameters as functions
of dimensionless distance, $\zeta$, in the small signal limit.}
\label{fig_radparms}
\end{figure}

The set of equations, eq.(\ref{eq_slicksix}), are soluble analytically. The first step
to a solution is to add eq.(\ref{eq_slicksix_i}) to eq.(\ref{eq_slicksix_q}): the right-hand
side of the combined equation is zero, so that $i_k + q_k$ is a constant, and under our
assumed background conditions,
\begin{equation}
i_k(\zeta) + q_k(\zeta) = 1 .
\label{eq_sol0}
\end{equation}
A second useful summation is to add eq.(\ref{eq_slicksix_g}) to eq.(\ref{eq_slicksix_h}), and
then to add the result to $\sqrt{2}$ times eq.(\ref{eq_slicksix_v}). The result is the
differential equation
\begin{equation}
d/d\zeta (\sqrt{2} v_k + g_k + h_k) = 2 (\sqrt{2} v_k + g_k + h_k) .
\label{eq_solbulge}
\end{equation}
If we set $\sigma_k = \sqrt{2} v_k + g_k + h_k$, eq.(\ref{eq_solbulge}) has the solution
$\sigma(\zeta) = \sigma (0) e^{2\zeta}$, but since $\sigma (0) = 0$ under our background
conditions, $\sigma (\zeta)$ is also zero for any larger distance. We therefore require that
\begin{equation}
g_k(\zeta) + h_k(\zeta) = -\sqrt{2} v_k(\zeta) .
\label{eq_sol1}
\end{equation}
We use eq.(\ref{eq_sol0}) and eq.(\ref{eq_sol1}) to eliminate $q_k$ and $v_k$ from eq.(\ref{eq_slicksix}).
Subtraction of the resulting equation in $h_k$ from its counterpart in $g_k$ then yields
\begin{equation}
d/d\zeta (g_k - h_k) = 2\sqrt{2} [i_k + j_k + (g_k - h_k)/\sqrt{2} - 1/2] ,
\label{eq_ykdef}
\end{equation}
and, as $g_k$ and $h_k$ appear only as the combination $g_k - h_k$ in the remaining equations, we
may introduce the new variable $y_k = g_k - h_k$, and write the three remaining equations as
\begin{subequations} \label{eq_last3}
\begin{align}
dj_k/d\zeta &= 2j_k + y_k/\sqrt{2} \label{eq_last3_j} \\
dy_k/d\zeta &= 2\sqrt{2} (i_k + j_k + y_k/\sqrt{2} - 1/2) \label{eq_last3_y} \\
di_k/d\zeta &= 2i_k + y_k/\sqrt{2} - 1  \label{eq_last3_i} .
\end{align}
\end{subequations}
We now let $f_k = i_k + j_k$, and add eq.(\ref{eq_last3_j}) to eq.(\ref{eq_last3_i}), leaving
the pair of equations,
\begin{subequations} \label{eq_pair}
\begin{align}
dy_k/d\zeta &= 2\sqrt{2} (f_k + y_k/\sqrt{2} - 1/2) \label{eq_pair_y} \\
df_k/d\zeta &= 2f_k + \sqrt{2} y_k - 1  \label{eq_last3_f} .
\end{align}
\end{subequations}
Reduction to a single equation is achieved by multiplying eq.(\ref{eq_last3_f}) by $\sqrt{2}$
and adding the result to eq.(\ref{eq_last3_y}). If we define $l_k = y_k + \sqrt{2} f_k$, the
final equation, in standard form as a first-order linear ODE, is
\begin{equation}
dl_k/d\zeta - 4 l_k = - 2\sqrt{2} .
\label{eq_leq}
\end{equation}
Equation~\ref{eq_leq} may be solved by standard methods, and it is then strightforward
to work back through the sequence of intermediate variables, and the constraints from
eq.(\ref{eq_sol0}) and eq.(\ref{eq_sol1}) to the solution of eq.(\ref{eq_slicksix}):
\begin{subequations} \label{eq_solution}
\begin{align}
i_k(\zeta) &= (e^{4\zeta} + 2 e^{2\zeta} + 5)/8 \label{eq_solution_i} \\
j_k(\zeta) &= (e^{4\zeta} - 2 e^{2\zeta} + 1)/8 \label{eq_solution_j} \\
q_k(\zeta) &= (3 - e^{4\zeta} - 2 e^{2\zeta})/8 \label{eq_solution_q} \\
g_k(\zeta) &= (e^{4\zeta} - 1)/(4\sqrt{2}) , \label{eq_solution_g}
\end{align}
\end{subequations}
together with the subsidary relations, $h_k (\zeta) = - g_k (\zeta)$ and, consequently
$v_k(\zeta)=0$. By inspection of eq.(\ref{eq_solution}), we can see that all the Stokes
and OAM parameters at moderate signal strengths (much greater than background, but not
saturating) tend to a rising exponential of the form $e^{4\zeta}$, so the OAM parameters
are likely to follow the polarization parameters to achieve levels such that
$j_k/i_k$ and $g_k/(\sqrt{2}i_k)$ tend to 1. The functions are plotted in the small
signal limit in Figure~\ref{fig_radparms}.

\subsection{Numerical Experiment}
\label{ss:numerics}

The analytic solutions in Section~\ref{ss:restricted} have demonstrated a non-trivial
coupling of the $z$-component of the electric field to the conventional Stokes parameters.
However, these results do not demonstrate a typical OAM pattern. In this section, we
consider the same quadrupole magnetic field structure, but sample the field at many
radii and azimuthal angles, $\phi$. The analytical solutions above may be considered
a single spatial sample from this array of points.

Additional geometrical considerations now need to be considered. The magnetic field magnitude now
increases outwards from the orign, where it is zero and the overlap of the $\pi$ and $\sigma$
responses is complete. There will therefore be no polarization or OAM generated at the origin.
As the radius is increased, the Zeeman splitting will rise, and the centre of the $\sigma^+$
response will move red-ward towards our chosen frequency, $\varpi_k$. As the radius continues
to increase, $\varpi_k$ will sample a decreasing wing of the Gaussian response. Although the
magnetic field intensity continues to rise, the exponential drop in the response will ensure
that, beyond a certain radius, there will be negligible amplification at $\varpi_k$. Rather
arbitrarily, we will set $\varpi_k$ to be -3 Doppler full widths ($k=-3$) to the red of the pattern
line centre, $\omega_0$. This value ensures a clean separation between the $\pi$ and $\sigma$
line shapes.

Our definition of $\zeta$ in this section differs slightly from that in Section~\ref{ss:restricted}.
Here, we reduce eq.(\ref{eq_stokesgold}) and eq.(\ref{eq_parmgold}) to dimensionless form by
dividing all equations by the group $D_{0,0}=|d_0|^2 P_0 \phi (0)/ \Gamma_0$: we have assumed
$M=0$ and symmetric pumping $P_M^+=P_M^-=P_M^0=P_0$, and similarly for the dipoles and
loss-rates. As a consequence, dimensionless inversion expressions, such as $R_{0,k}$ and $\Pi_{0,k}$ reduce
simply to differences of various Gaussians, centered on the associated molecular responses.
For example,
\begin{equation}
R_{0,k} = e^{-[(\varpi_k - \Delta \omega_0^+ )/\Delta \omega_D ]^2 } - e^{ -[(\varpi_k - \Delta \omega_0^-)/\Delta \omega_D ]^2 } ,
\label{eq_Rexample}
\end{equation}
where $\Delta \omega_D = w \omega_0 /c$, and $w$ is in turn defined in eq.(\ref{eq_gwidth}).
All radiation parameters are scaled by the level of the input seed radiation in Stokes-$I$; the
other radiation parameters all have a background level of $0$.
The dimensionless evolution equations were solved using a Runge-Kutta fourth-order method with
adaptive step-size control, maintained by a fifth-order accuracy checker
\citep{1992nrfa.book.....P}. The results discussed below were drawn from a run
with an integration accuracy of $2 \times 10^{-7}$.

\begin{figure*}
  \includegraphics[width=175mm]{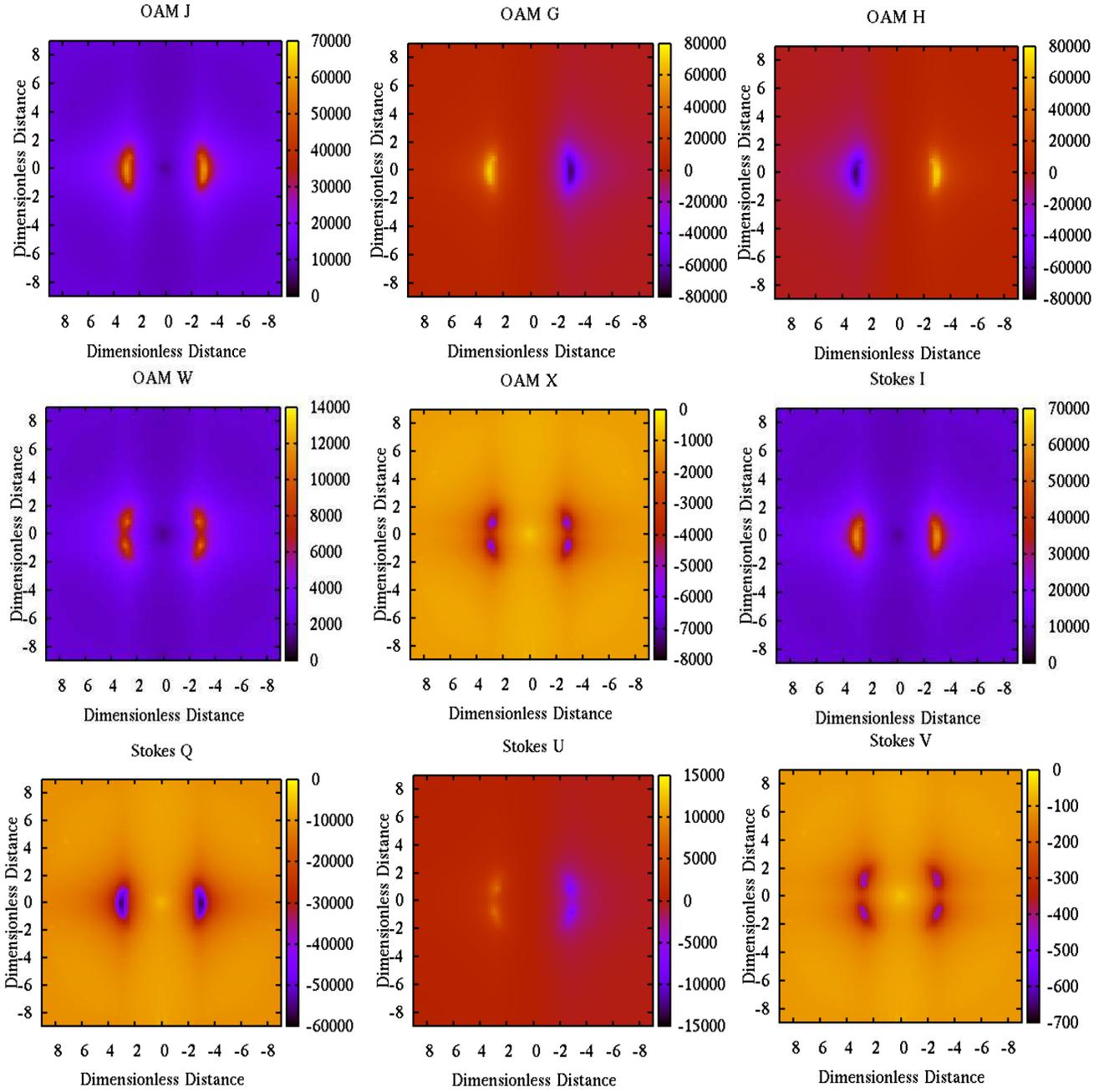}
  \caption{The emergent radiation parameters of a single Fourier component in the sky plane
after propagation along the $z$-axis over a
dimensionless distance of $\zeta = 4.0$. Each sub-figure has its own colour scale, shown to
the right of the diagram. Figures on this scale are multiples of the seed value of Stokes-$I$.
The $x$ and $y$ axes are drawn to conform to the IAU convention: with
radiation approaching, $x$ increases up the page, and $y$, to the left. The $x$ and $y$ axes
are in Doppler or Zeeman units, so a distance of three units from the origin produces a
magnetic field capable of shifting the response of a $\sigma$-transition by three Doppler
widths from the frequency of the central $\pi$-component.}
\label{fig_composite}
\end{figure*}

In Figure~\ref{fig_composite}, we plot the emergent radiation parameters as a function of 
distance from the origin (position of zero magnetic field) on Cartesian axes. 
These axes conform to the IAU standard, as discussed in
Section~\ref{axes}, so that the $x$-axis points vertically up the page (North) and the $y$-axis
increases to the left. Distances are in magnetic field strength units, as required to yield a certain
Zeeman splitting in Doppler widths. The azimuthal angle, $\phi$ is
measured anticlockwise from its zero position at North.
 The $z$-axis, along which the radiation propagated, points out of the
page towards the observer, and all parameters have been amplified over the same
dimensionless distance of $\zeta = 4.0$, a value that corresponds to moderate amplification:
peak values of Stokes-$I$ are vastly greater than the background, but still low enough that
saturation is negligible for a realistic OH maser that is amplifying a CMB background.

In all cases, the radial structure of the radiation pattern is controlled by a combination
of the magnetic field strength (which increases outwards) and the Doppler response of the
molecules, which may be considered Gaussian. We see the brightest emission close to a
radius of $3$, where the magnetically shifted response in the $\sigma^+$ part of the Zeeman
pattern lies closest to our chosen Fourier component with $k=-3$. Different Fourier
components will exhibit different patterns. Choosing $k=+3$ would produce a very similar
radial pattern, but with the signs of some parameters reversed.

The angular distributions are determined partly by the angular functions of $\phi'$ that
appear in eq.(\ref{eq_parmgold}) and eq.(\ref{eq_stokesgold}) (variously sines and cosines
of either $\phi'$ or $2 \phi'$) and partly by the exponential nature of the
maser amplification. In particular, all OAM parameters are zero, as are
Stokes-$U$ and $V$, at $\phi = \phi' = 0$, the
standard reduction to the case of a uniform magnetic field. This confirms the
hypothesis that OAM cannot be generated from a uniform field, an outcome that is
not obvious by inspection of eq.(\ref{eq_parmgold}).

A difference between Stokes-$I$ and the OAM-$J$ parameter that is not immediately
apparent in Figure~\ref{fig_composite} is that $I$ is not zero at $\phi=0$: instead, it
has a value of $1.45 \times 10^3$, compared to its peak of $1.05 \times 10^6$ at
$\phi = \pi/2$ and $3\pi/2$. The maxima of $I$ and $J$ are very similar, indicating
that the fractional level of OAM, like polarization, may become very large without
invoking saturation. The only other parameter that is non-zero at $\phi=0$ (and $\pi$)
is Stokes-$Q$: it is {\em positive} at this angle, with a value of $1.45 \times 10^3$,
before giving way to negative values of much larger magnitude at angles closer to the $y$-axis.
Physically, the larger values of Stokes-$I$ near $\phi=\pi/2$, compared to $\phi=0$, probably result
from the radiation at the former position coupling to all three Cartesian components
of the molecular dipole; at the latter position, corresponding to the polarization-only
case, radiation can couple only to the dipole components in the $xy$-plane.

There is a general hierarchy of (absolute) peak values: levels of approximately $10^6$ are
achieved by the Stokes parameters $I$ and $Q$, by the $z$-component intensity, $J$, and
the two symmetric coupling parameters $G$ and $H$. A second set, comprising the asymmetric
coupling parameters $W$ and $X$, together with Stokes $U$, achieve peaks of order
one order of magnitude lower $\sim 10^5$. Finally Stokes-$V$ is very weakly amplified,
always negative, and has an absolute peak of $\sim 700$. It is probably also worth
noting that whilst non-zero values of $W$ are exclusively positive, $X$ it not
completely negative, achieving a positive peak of approximately 680.

The exponential nature of maser amplification introduces a number of feaures that
are not typical of radiation beams with OAM in the laboratory. For example, a 
laboratory beam in an L-G mode with azimuthal order $l=2$ might be expected to
have an annular intensity pattern, and a $\sin 2\phi$ structure in phase. It
should be pointed out here that the intensity, as represented by Stokes-$i$ in
Figure~\ref{fig_composite}, is annular, but amplified interaction with other
Stokes and OAM parameters have introduced an additional very strong angular
structure. Parameters more sensitive to the phase, such as $G,H,W,X$ show
a pattern that is close to $\sin \phi$ (for $G$,$H$) and $\sin 2\phi$ (in $W$ and $X$) at
low amplification, but these patterns again become distorted by exponential 
growth. Peaks in the weaker parameters, for example $W$ and $X$, at odd multiples of $\pi/4$
migrate with amplification towards the peaks of the stronger parameters.

Lack of linearly polarized Zeeman triplet patterns is a well-known observational 
feature of OH maser sources. Such patterns would logically arise from the
propagation of maser radiation perpendicular to magnetic field lines, but
appear rare compared to the Zeeman pairs of opposite-handed circular or
elliptically polarized components generated by propagation that is close to
aligned with the magnetic field. Many reasons for the lack of triplets have
been suggested, including MHD turbulence \citep{1995ApJ...438..763G}, Faraday depolarization within
the source \citep{1973ApJ...179..111G} and preferential beaming along 
magnetic field lines \citep{1994A&A...292..693G}.
Another possibility, arising from the present work, is that linearly-polarized
masers may also possess high degrees of OAM, and may therefore be
invisible to conventional radio detection equipment. It is certainly most
unlikely that magnetic fields perpendicular to the line of sight will be
totally uniform, but they do seem to be ordered on the scale of the source, down
to a clustering scale of order 70\,AU. For an OAM scheme to work, magnetic
fields would need to be non-uniform on the scale of individual VLBI maser
spots: perhaps 10\,AU or even less, and evidence that the sky-component of
the magnetic field tends to be aligned with the long axis of an individual
spot \citep{2006ApJS..164...99F} suggests some ordering of the field even at
this scale. Nevertheless, while the arrangement used in the present work is
highly idealized, a significant yield of radiation with OAM from at
least some OH masers seems likely.

The radiation patterns produced as a function of radius and
angle in the present work do not correspond to any single 
Laguerre-Gaussian mode. However, they do reveal the underlying
symmetry of the magnetic field, and the spectrum of modes present
is therefore likely to yield information about the structure
of the magnetic field on scales smaller than the size of
the maser spot itself. By contrast, polarization, assuming
a uniform field in each spot, can only tell us about the
variation of the magnetic field over an area of sky
containing many spots. The fraction of radiation converted
to OAM also helps us to reconstruct the magnetic field in
3-D, since a field parallel to the line of sight produces
no OAM (if the background radiation has none), whilst increasing
amounts of OAM result as $\theta$ is increased towards $\pi /2$.

\section{Conclusions}
\label{conclusion}

We have extended the standard theory of propagation of polarized astrophysical maser radiation
to the case of a non-uniform magnetic field, allowing for the presence of a component of
the electric field of the radiation in the propagation direction. A set of equations for the
evolution of the complex amplitudes of the radiation has been derived in the time domain, and
converted to the frequency domain, where we consider many finite-width Fourier components of the radiation field,
collectively extending across the full Zeeman pattern. A classical reduction of the frequency-domain
equations leads to a set of nine coupled differential equations for the distance evolution of
the standard Stokes parameters and five additional parameters that represent a coupling
of the $z$-component of the electric field to the usual $x$ and $y$ components. These latter
five parameters may represent radiation with orbital angular momentum (OAM).

There is a non-trivial coupling between the response of Zeeman-split molecules in
a non-uniform magnetic field, and the electric field of radiation in the direction
of propagation, and this coupling is strongest when the magnetic field is perpendicular
to the propagation direction. Although it is not obvious from the evolution equations,
the standard reduction to a uniform field (setting the azimuthal angle $\phi'=0$) results
in radiation that may have polarization, but no OAM. This result was confirmed in
Section~\ref{ss:numerics}. The OAM coupling is also most effective when, for a selected
Fourier component or frequency, there is a large difference between the inversions in
the various $\pi$- and $\sigma$-transitions of the Zeeman pattern. If we assume symmetric
pumping, this implies that OAM generation will be significantly more efficient in
molecules with a large Zeeman splitting (for example OH) than in closed-shell species
(for example SiO, water and methanol).

A restricted analytical solution demonstrates that the coupling of the parameters
representing OAM to the Stokes parameters, particularly Stokes-$I$, is non-trivial.
For a suitable non-uniform magnetic field, in this case an ideal quadrupole, OAM
can be generated from seed radiation without OAM or polarization, just as polarization
may be so generated in a uniform field. Levels of OAM may grow large without the
need for maser saturation.

A trial computational solution of the governing equations shows that in an intermediate
amplification regime (intensity vastly greater than the background, but not saturating)
OAM parameters may become large, at least with a non-uniform magnetic field of rather
ideal structure. Maser radiation propagated perpendicular to the magnetic field may
therefore evolve OAM fractions approaching 100 per cent. There is a hierarchy of
amplification levels with Stokes-$I$ and the OAM parameters $J$,$G$ and $H$ as the
strongest set, followed by $W$, $X$ and Stokes $U$, and finally a weakly amplified Stokes-$V$.
Radiation patterns depart from usual OAM expectations owing to the exponential
amplification of angular structure.

OAM conversion may partially account for the loss of linearly-polarized OH masers
if non-uniform magnetic fields are common at the scales typical of resolved
VLBI maser spots. Additional diagnostic value of OAM radiation, in the
context of the present work, is discussed at the end of
Section~\ref{ss:numerics}.

\section*{Acknowledgments}

Computations were carried out on the Legion supercomputer at the
HiPerSPACE Computing Centre, University College London, which
is funded by the UK Science and Technology Facilities Council (STFC).

\bibliographystyle{mn2e}
\bibliography{poam2}



\end{document}